\documentclass[journal]{IEEEtran}

\usepackage[nosort]{cite}

\ifCLASSINFOpdf
\else
\fi

\usepackage{amsmath}
\usepackage{algorithmic}
\usepackage{array}
\usepackage{stfloats}
\usepackage{xcolor}
\usepackage{indentfirst}
\usepackage{epsfig,setspace}
\usepackage{float,algorithm}
\usepackage{bm}
\usepackage{multirow, hhline}
\usepackage{makecell}
\usepackage{amssymb}
\usepackage{pifont}
\usepackage{arydshln}
\usepackage{graphicx}
\usepackage{booktabs}
\newcommand{\cmark}{\ding{51}}
\newcommand{\xmark}{\ding{55}}
\newcommand{\tabincell}[2]{\begin{tabular}{@{}#1@{}}#2\end{tabular}}
\usepackage{subfig}
\usepackage{caption}
\usepackage{amsmath}

\begin{document}
\bstctlcite{IEEEexample:BSTcontrol}

\title{Speaker Adaptation Using Spectro-Temporal Deep Features for Dysarthric and Elderly Speech Recognition}

\author{Mengzhe Geng, Xurong Xie, Zi Ye, Tianzi Wang, Guinan Li, Shujie Hu, \\ Xunying Liu,~\IEEEmembership{Member,~IEEE}, Helen Meng,~\IEEEmembership{Fellow,~IEEE}
    \thanks{Mengzhe Geng,  Zi Ye, Tianzi Wang, Guinan Li and Shujie Hu are with the Chinese University of Hong Kong, China (email: \{mzgeng,zye,twang,gnli,sjhu\}@se.cuhk.edu.hk).\\
    \indent Xurong Xie is with Institute of Software, Chinese Academy of Sciences, Beijing, China (email: xurong@iscas.ac.cn).\\
    \indent Xunying Liu is with the Chinese University of Hong Kong, China and the corresponding author (email: xyliu@se.cuhk.edu.hk).\\
    \indent Helen Meng is with the Chinese University of Hong Kong, China (email: hmmeng@se.cuhk.edu.hk).}
}

\markboth{JOURNAL OF \LaTeX\ CLASS FILES, VOL. 14, NO. 8, AUGUST 2015}%
{Shell \MakeLowercase{\textit{et al.}}: nothing goes here}

\maketitle

\begin{abstract}
Despite the rapid progress of automatic speech recognition (ASR) technologies targeting normal speech in recent decades, accurate recognition of dysarthric and elderly speech remains highly challenging tasks to date. Sources of heterogeneity commonly found in normal speech including accent or gender, when further compounded with the variability over age and speech pathology severity level, create large diversity among speakers. To this end, speaker adaptation techniques play a key role in personalization of ASR systems for such users. Motivated by the spectro-temporal level differences between dysarthric, elderly and normal speech that systematically manifest in articulatory imprecision, decreased volume and clarity, slower speaking rates and increased dysfluencies, novel spectro-temporal subspace basis deep embedding features derived using SVD speech spectrum decomposition are proposed in this paper to facilitate auxiliary feature based speaker adaptation of state-of-the-art hybrid DNN/TDNN and end-to-end Conformer speech recognition systems. Experiments were conducted on four tasks: the English UASpeech and TORGO dysarthric speech corpora; the English DementiaBank Pitt and Cantonese JCCOCC MoCA elderly speech datasets. The proposed spectro-temporal deep feature adapted systems outperformed baseline i-Vector and x-Vector adaptation by up to 2.63\% absolute (8.63\% relative) reduction in word error rate (WER). Consistent performance improvements were retained after model based speaker adaptation using learning hidden unit contributions (LHUC) was further applied. The best speaker adapted system using the proposed spectral basis embedding features produced the lowest published WER of 25.05\% on the UASpeech test set of 16 dysarthric speakers. 
\end{abstract}


%
\IEEEpeerreviewmaketitle

\section{Introduction}
\label{sec-intro}
\IEEEPARstart{D}{espite} the rapid progress of automatic speech recognition (ASR) techonologies targeting normal speech in recent decades~\cite{bahl1986maximum,graves2013speech,peddinti2015time,chan2016listen,wang2020transformer,gulati2020conformer,hu2021neural,hu2021bayesian}, accurate recognition of dysarthric and elderly speech remains highly challenging tasks to date~\cite{christensen2012comparative,christensen2013combining,sehgal2015model,yu2018development,hu2019cuhk,liu2020exploiting,ye2021development,jin2021adversarial}. Dysarthria is caused by a range of speech motor control conditions including cerebral palsy, amyotrophic lateral sclerosis, stroke and traumatic brain injuries~\cite{whitehill2000speech,makkonen2018speech,scott1983speech,jerntorp1992stroke,lanier2010speech}. In a wider context, speech and language impairments are also commonly found among older adults experiencing natural ageing and neurocognitive disorders, for example, Alzheimer’s disease~\cite{fraser2016linguistic, wiley2021alzheimer}. People with speech disorders often experience co-occurring physical disabilities and mobility limitations. Their difficulty in using keyboard, mouse and touch screen based user interfaces makes voice based assistive technologies more natural alternatives~\cite{hux2000accuracy,young2010difficulties} even though speech quality is degraded. To this end, in recent years there has been increasing interest in developing ASR technologies that are suitable for dysarthric and elderly speech~\cite{vipperla2010ageing,christensen2013combining,rudzicz2014speech,zhou2016speech,vachhani2017deep,kim2018dysarthric,konig2018fully,joy2018improving,toth2018speech,yu2018development,liu2019exploiting,mirheidari2019dementia,Shor2019,geng2020investigation,lin2020staged,kodrasi2020spectro,xiong2020source,liu2021recent,jin2021adversarial,hu2021bayesian,xie2021variational,takashima2020two,hermann2020dysarthric,wang2021improved,wang2021study,macdonald2021disordered,green2021automatic,ye2021development,pan2021using,geng2021spectro}.

Dysarthric and elderly speech bring challenges on all fronts to current deep learning based automatic speech recognition technologies predominantly targeting normal speech recorded from healthy, non-aged users. In addition to the scarcity of such data, their large mismatch against healthy speech and the difficulty in collecting them on a large scale from impaired and elderly speakers due to mobility issues, the need of modelling the prominent heterogeneity among speakers is particularly salient. Sources of variability commonly found in normal speech including accent or gender, when further compounded with those over age and speech pathology severity, create large diversity among dysarthric and elderly speakers ~\cite{kodrasi2020spectro,smith1987temporal}. The deficient abilities in controlling the articulators and muscles responsible for speech production lead to abnormalities in dysarthric and elderly speech manifested across many fronts including articulatory imprecision, decreased volume and clarity, increased dysfluencies, changes in pitch and slower speaking rate~\cite{kent2000dysarthrias}. In addition, the temporal or spectral perturbation based data augmentation techniques~\cite{vachhani2018data,xiong2019phonetic,geng2020investigation} that are widely used in current systems to circumvent data scarcity further contribute to speaker-level variability. To this end, speaker adaptation techniques play a key role in personalization of ASR systems for such users. Separate reviews over conventional speaker adaptation techniques developed for normal speech and those for dysarthric or elderly speech are presented in the following Section \ref{sec-intro-normal} and \ref{sec-intro-dys-elderly}.

\subsection{Speaker Adaptation for Normal Speech}
\label{sec-intro-normal}

Speaker adaptation techniques adopted by current deep neural networks (DNNs) based ASR systems targeting normal speech can be divided into three major categories: 1) auxiliary speaker embedding feature based methods that represent speaker dependent (SD) features via compact vectors~\cite{abdel2013fast,saon2013speaker,senior2014improving,huang2015investigation}, 2) feature transformation based approaches that produce speaker independent (SI) canonical features at the acoustic front-ends~\cite{digalakis1995speaker,lee1996speaker,gales1998maximum,uebel1999investigation,seide2011feature} and 3) model based adaptation techniques that compensate the speaker-level variability by often incorporating additional SD transformations that are applied to DNN parameters or hidden layer outputs~\cite{neto1995speaker,anastasakos1996compact,gemello2007linear,li2010comparison,swietojanski2016learning,zhang2016dnn}. 

In the auxiliary speaker embedding feature based approaches, speaker dependent (SD) features such as speaker codes~\cite{abdel2013fast} and i-Vectors~\cite{saon2013speaker,senior2014improving} are concatenated with acoustic features to facilitate speaker adaptation during both ASR system training and evaluation. The estimation of SD auxiliary features can be performed independently of the remaining recognition system components. For example, i-Vectors~\cite{saon2013speaker,senior2014improving} are learned from Gaussian mixture model (GMM) based universal background models (UBMs). The SD auxiliary features can also be jointly estimated with the back-end acoustic models, for example, via an alternating update between them and the remaining SI DNN parameters in speaker codes~\cite{abdel2013fast}. Auxiliary feature based speaker adaptation methods benefit from both their low complexity in terms of the small number of SD feature parameters to be estimated, and their flexibility allowing them to be incorporated into a wide range of ASR systems including both hybrid DNN-HMM systems and recent end-to-end approaches~\cite{tuske2021limit}.

In feature transformation based speaker adaptation, feature transforms are applied to acoustic front-ends to produce canonical, speaker invariant inputs. These are then fed into the back-end DNN based ASR systems to model the remaining sources of variability, for example, phonetic and phonological context dependency in speech. Feature-space maximum likelihood linear regression (f-MLLR) transforms~\cite{seide2011feature} estimated at speaker-level from GMM-HMM based ASR systems~\cite{digalakis1995speaker,gales1998maximum} are commonly used. In order to account for the vocal tract length differences between speakers, physiologically motivated vocal tract length normalization (VTLN) can also be used as feature transformation~\cite{lee1996speaker,uebel1999investigation}. Speaker-level VTLN normalized features can be obtained using either piecewise linear frequency warping factors directly applied to the spectrum, or affine linear transformations akin to f-MLLR.

In model based adaptation approaches, separately designed speaker-dependent DNN model parameters are used to account for speaker-level variability. In order to ensure good generalization and reduce the risk of overfitting to limited speaker-level data, a particular focus of prior researches has been on deriving compact forms of SD parameter representations. These are largely based on linear transforms that are incorporated into various parts of DNN acoustic models. These include the use of SD linear input networks (LIN)~\cite{digalakis1995speaker,li2010comparison}, linear output networks (LON)~\cite{lanier2010speech}, linear hidden networks (LHN)~\cite{gemello2007linear}, learning hidden unit contributions (LHUC)~\cite{swietojanski2014learning,swietojanski2016learning,zhang2016dnn}, parameterized activation functions (PAct)~\cite{zhang2015parameterised,zhang2016dnn}, speaker-independent (SI) and SD factored affine transformations~\cite{zhao2016low}, and adaptive interpolation of outputs of basis sub-networks~\cite{wu2015multi,tan2015cluster}. In addition to only modelling speaker-level variability in the test data during recognition, the estimation of SD parameters in both the system training and evaluation stages leads to more powerful speaker adaptive training (SAT)~\cite{anastasakos1996compact} approaches, allowing a joint optimization of both the SD and SI parameters during system training. 

\subsection{Speaker Adaptation for Dysarthric and Elderly Speech}
\label{sec-intro-dys-elderly}
In contrast, only limited research on speaker adaptation techniques targeting dysarthric and elderly speech recognition has been conducted so far. Earlier works in this direction were mainly conducted in the context of traditional GMM-HMM acoustic models. The application of maximum likelihood linear regression (MLLR) and maximum a posterior (MAP) adaptation to such systems were investigated in~\cite{baba2002elderly, mengistu2011adapting,christensen2012comparative,kim2013dysarthric}. MLLR was further combined with MAP adaptation in speaker adaptive training (SAT) of SI GMM-HMM in~\cite{sehgal2015model}. F-MLLR based SAT training of GMM-HMM systems was investigated in~\cite{bhat2016recognition}. Regularized speaker adaptation using Kullback-Leibler (KL) divergence was studied for GMM-HMM systems in~\cite{kim2017regularized}.

More recent researches applied model adaptation techniques to a range of state-of-the-art DNN based dysarthric and elderly speech recognition systems. Normal to dysarthric speech domain adaptation approaches using direct parameter fine-tuning were investigated in both lattice-free maximum mutual information (LF-MMI) trained time delay neural network (TDNN)~\cite{xiong2020source,takashima2020two} based hybrid ASR systems and end-to-end recurrent neural network transducer (RNN-T)~\cite{Shor2019,green2021automatic} systems. In order to mitigate the risk of overfitting to limited speaker-level data during model based adaptation, more compact learning hidden unit contributions (LHUC) based dysarthric speaker adaptation was studied in~\cite{yu2018development,geng2020investigation,liu2021recent} while Bayesian learning inspired domain speaker adaptation approaches have also been proposed in~\cite{deng2021bayesian}. 

One main issue associated with previous researches on dysarthric and elderly speaker adaptation is that the systematic, fine-grained speaker-level diversity attributed to speech impairment severity and aging is not considered. Such diversity systematically manifests itself in a range of spectro-temporal characteristics including articulatory imprecision, decreased volume and clarity, breathy and hoarse voice, increased dysfluencies as well as slower speaking rate.

In order to address this issue, novel deep spectro-temporal embedding features are proposed in this paper to facilitate auxiliary speaker embedding feature based adaptation for dysarthric and elderly speech recognition. Spectral and temporal basis vectors derived by singular value decomposition (SVD) of dysarthric or elderly speech spectra were used to structurally and intuitively represent the spectro-temporal key attributes found in such data, for example, an overall decrease in speaking rate and speech volume as well as changes in the spectral envelope. These two sets of basis vectors were then used to construct DNN based speech pathology severity or age classifiers. More compact, lower dimensional speaker specific spectral and temporal embedding features were then extracted from the resulting DNN classifiers' bottleneck layers, before being further utilized as auxiliary speaker embedding features to adapt start-of-the-art hybrid DNN~\cite{liu2021recent}, hybrid TDNN~\cite{peddinti2015time} and end-to-end (E2E) Conformer~\cite{gulati2020conformer} ASR systems. 

Experiments were conducted on four tasks: the English UASpeech~\cite{kim2008dysarthric} and TORGO~\cite{rudzicz2012torgo} dysarthric speech corpora; the English DementiaBank Pitt~\cite{becker1994natural} and Cantonese JCCOCC MoCA~\cite{xu2021speaker} elderly speech datasets. Among these, UASpeech is by far the largest available and widely used dysarthric speech database, while DementiaBank Pitt is the largest publicly available elderly speech corpus. The proposed spectro-temporal deep feature adapted systems outperformed baseline i-Vector~\cite{saon2013speaker} and x-Vector~\cite{snyder2018x} adapted systems by up to $2.63\%$ absolute ($8.63\%$ relative) reduction in word error rate (WER). Consistent performance improvements were retained after model based speaker adaptation using learning hidden unit contributions (LHUC) was further applied. 
The best speaker adapted system using the proposed spectral basis embedding features produced the lowest published WER of $25.05\%$ on the UASpeech test set of $16$ dysarthric speakers. Speech pathology severity and age prediction performance as well as further visualization using t-distributed stochastic neighbor embedding (t-SNE)~\cite{van2008visualizing} indicate that our proposed spectro-temporal deep features can more effectively learn the speaker-level variability attributed to speech impairment severity and age than conventional i-Vector~\cite{saon2013speaker} and x-Vector~\cite{snyder2018x}. The main contributions of this paper are summarized below: 

1) To the best of our knowledge, this paper presents the first use of spectro-temporal deep embedding features to facilitate speaker adaptation for dysarthric and elderly speech recognition. In contrast, there were no prior researches published to date on auxiliary features based speaker adaptation targeting such data. Existing speaker adaptation methods for dysarthric and elderly speech use mainly  model based approaches~\cite{baba2002elderly,mengistu2011adapting,christensen2012comparative,kim2013dysarthric,bhat2016recognition,sehgal2015model,yu2018development,Shor2019,xiong2020source,geng2020investigation}. Speaker embedding features were previously only studied for speech impairment assessment~\cite{an2015automatic,garcia2018multimodal,janbakhshi2020subspace}.

2) The proposed spectro-temporal deep features are inspired and intuitively related to the latent variability of dysarthric and elderly speech. The spectral basis embedding features are designed to learn characteristics such as volume reduction, changes of spectral envelope, imprecise articulation as well as breathy and hoarse voice, while the temporal basis embedding features to capture patterns such as increased dysfluencies and pauses. The resulting fine-grained, factorized representation of diverse impaired speech characteristics serves to facilitate more powerful personalized user adaptation for dysarthric and elderly speech recognition.

3) The proposed spectro-temporal deep feature adapted systems achieve statistically significant performance improvements over baseline i-Vector or x-Vector adapted hybrid DNN/TDNN and end-to-end (E2E) Conformer systems by up to $2.63\%$ absolute ($8.63\%$ relative) word error rate (WER) reduction on four dysarthric or elderly speech recognition tasks across two languages. These findings serve to demonstrate the efficacy and genericity of our proposed spectro-temporal deep features for dysarthric and elderly speaker adaptation. 

The rest of this paper is organized as follows. The derivation of spectro-temporal basis vectors using SVD speech spectrum decomposition is presented in Section~\ref{sec-decomposition}. The extraction of spectro-temporal deep embedding features and their incorporation into hybrid DNN/TDNN and end-to-end Conformer based ASR systems for speaker adaptation are proposed in Section~\ref{sec-feature}. A set of implementation issues affecting the learning of spectro-temporal deep embedding features are discussed in Section~\ref{sec-implementation}. Section~\ref{sec-experiment} presents the experimental results and analysis. Section~\ref{sec-conclusion} draws the conclusion and discusses possible future works.

\section{Speech Spectrum Subspace Decomposition}
\label{sec-decomposition}
Spectro-temporal subspace decomposition techniques provide a simple and intuitive solution to recover speech signals from noisy observations by modelling the combination between these two using a linear system~\cite{van1993subspace}. This linear system can then be solved using signal subspace decomposition schemes, for example, singular value decomposition (SVD)~\cite{van1993subspace,janbakhshi2020subspace} or non-negative matrix factorization (NMF) methods~\cite{lee1999learning, fevotte2009nonnegative,wang2012online}, both of which are performed on the time-frequency speech spectrum. 

\begin{figure*}[ht]
  \centering
    \subfloat[DYS v.s. CTL from UASpeech corpus]{\includegraphics[width=0.49\textwidth]{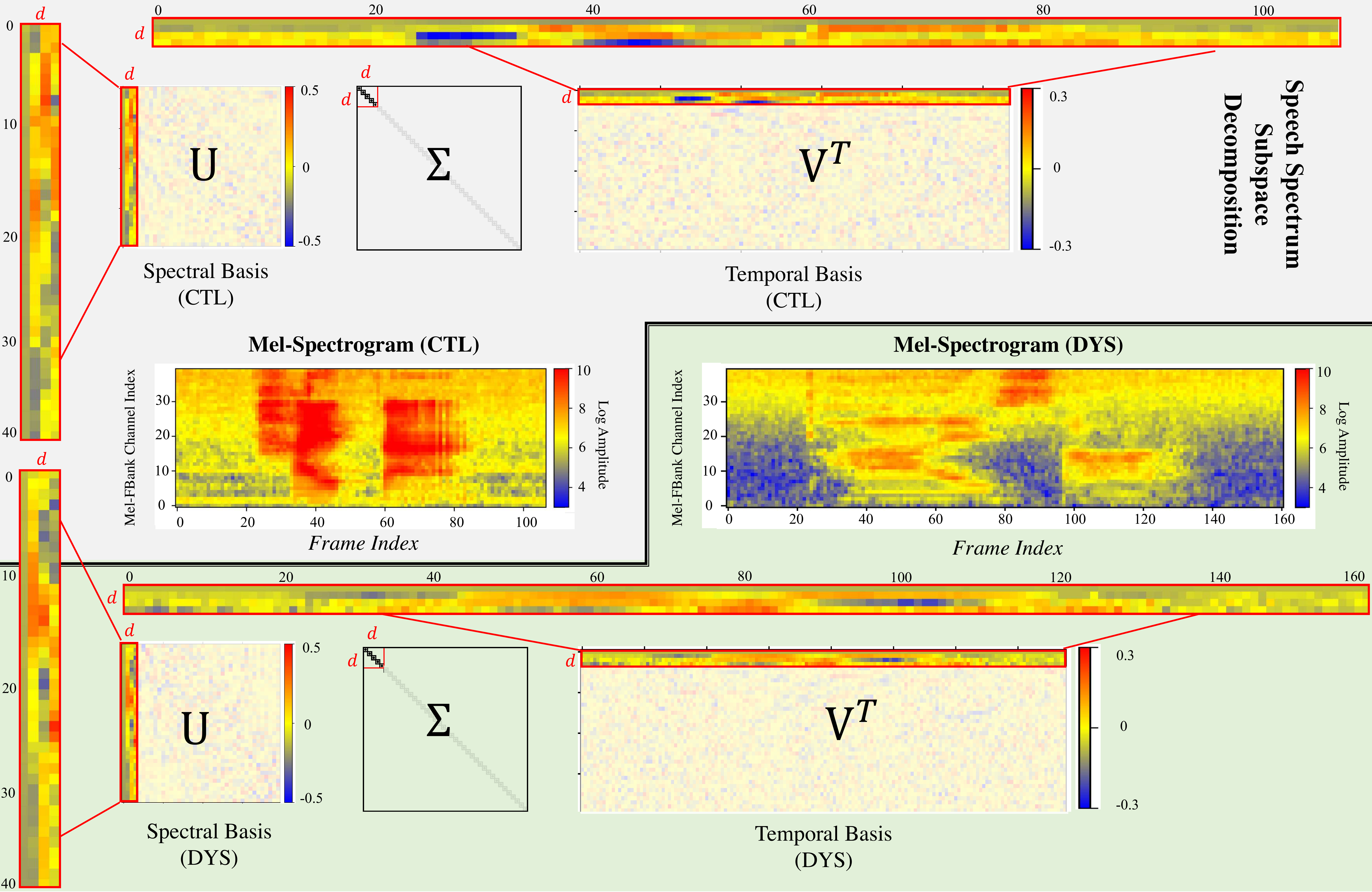}
    \label{fig:SVD-Example-UASpeech}}  
    \hfill
    \subfloat[PAR v.s. INV from DBANK corpus]{\includegraphics[width=0.49\textwidth]{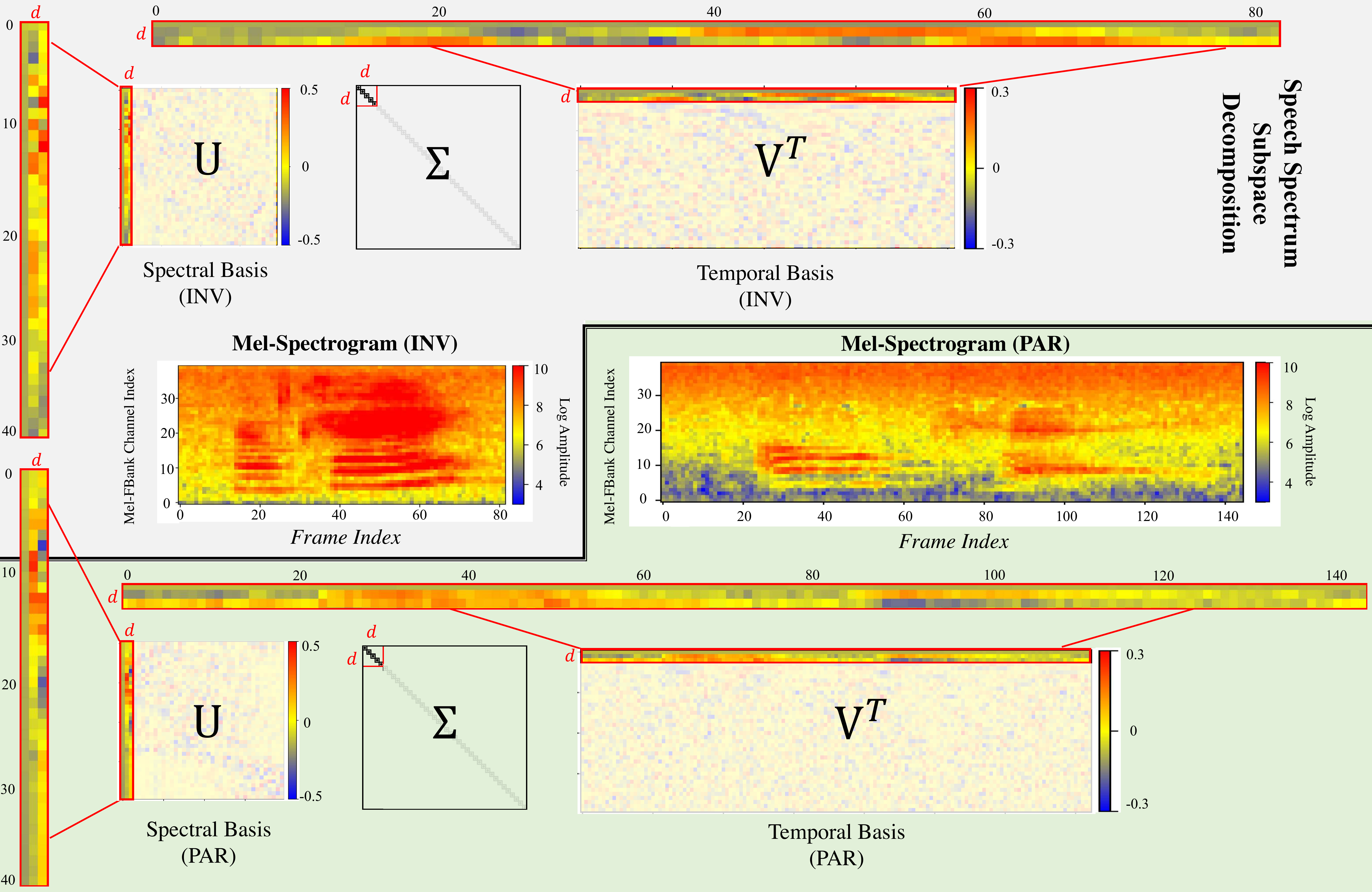}
    \label{fig:SVD-Example-DBANK}}
    \caption{Example subspace decomposition of mel-spectrogram of: (a) a pair of normal (CTL, left upper) and dysarthric (DYS, left lower) utterances of word ``python'' to obtain top $d=4$ spectral and temporal basis vectors (circled in red in $\mathbf{U}$ and $\mathbf{V^{\mathrm{T}}}$) of the UASpeech~\cite{kim2008dysarthric} corpus; and (b) a pair of non-aged clinical investigator (INV, right upper) and elderly participant (PAR, right lower) utterances of word “okay” to obtain top $d=3$ spectral and temporal basis vectors (circled in red in $\mathbf{U}$ and $\mathbf{V^{\mathrm{T}}}$) of the DementiaBank Pitt (DBANK)~\cite{becker1994natural} dataset.}
    \label{fig:SVD-Example}
\end{figure*}

An example SVD decomposition of a mel-scale filter-bank based log amplitude spectrum is shown in Fig.~\ref{fig:SVD-Example-UASpeech} and~\ref{fig:SVD-Example-DBANK}. Let $\mathbf{S_r}$ represent a $C \times T$ dimensional mel-scale spectrogram of utterance $r$ with $C$ filter-bank channels and $T$ frames. The SVD decomposition~\cite{van1993subspace} of $\mathbf{S_r}$ is given by:

\begin{equation}
    \mathbf{S_{r} = U_{r}{\Sigma_{r}}V_{r}^{\mathrm{T}}}
\end{equation}

where the set of column vectors of the $C \times C$ dimensional left singular $\mathbf{U_r}$ matrix and the row vectors of the $T \times T$ dimensional right singular $\mathbf{V_{r}^{\mathrm{T}}}$ matrix are the bases of the spectral and temporal subspaces respectively. Here $\mathbf{\Sigma_{r}}$ is a $C \times T$ rectangular diagonal matrix containing the singular values sorted in a descending order, which can be further absorbed into a multiplication with $\mathbf{V_{r}^{\mathrm{T}}}$ for simplicity. In order to obtain more compact representation of the two subspaces, a low-rank approximation~\cite{fevotte2009nonnegative} obtained by selecting the top-$d$ principal spectral and temporal basis vectors can be used. In this work, the number of principal components $d$ is empirically set to vary from $2$ to $10$. 

The SVD decomposition shown in Fig.~\ref{fig:SVD-Example} intuitively separates the speech spectrum into two sources of information that can be related to the underlying sources of variability in dysarthric and elderly speech: a) \textbf{time-invariant spectral subspaces} that can be associated with an average utterance-level description of dysarthric or elderly speakers' characteristics such as an overall reduction of speech volume, changes in the spectral envelope shape, weakened formats due to articulation imprecision as well as hoarseness and energy distribution anomaly across frequencies due to difficulty in breath control. For example, the comparison between the spectral basis vectors extracted from a pair of dysarthric and normal speech utterances of the identical content ``python'' in Fig.~\ref{fig:SVD-Example-UASpeech} shows that the dysarthric spectral basis vectors exhibit a pattern of energy distribution over mel-scale frequencies that differs from that obtained from the normal speech spectral bases. Similar trends can be found between the spectral basis vectors of non-aged and elderly speech utterances of the same word content ``okay'' shown in Fig.~\ref{fig:SVD-Example-DBANK}. b) \textbf{time-variant temporal subspaces} that are considered more related to sequence context dependent features such as decreased speaking rate as well as increased dysfluencies and pauses, for example, shown in the contrast between the temporal basis vectors separately extracted from normal and dysarthric speech in Fig.~\ref{fig:SVD-Example-UASpeech} and those from non-aged and elderly speech in Fig.~\ref{fig:SVD-Example-DBANK}, where the dimensionality of the temporal subspace captures the speaking rate and duration. 

SVD spectrum decomposition is performed in an unsupervised fashion. In common with other unsupervised feature decomposition methods such as NMF, it is theoretically non-trivial to produce a perfect disentanglement~\cite{locatello2020sober} between the time-invariant and variant speech characteristics encoded by the spectral and temporal basis vectors respectively, as both intuitively represent certain aspects of the underlying speaker variability associated with speech pathology severity and age.

For the speaker adaptation task considered in this paper, the ultimate objective is to obtain more discriminative feature representations to capture dysarthric and elderly speaker-level diversity attributed to speech impairment severity and age. To this end, further supervised learning of deep spectro-temporal embedding features is performed by constructing deep neural network based speech pathology severity or age classifiers taking the principal spectral or temporal basis vectors as their inputs. These are presented in the following Section~\ref{sec-feature}.

\section{Spectro-Temporal Deep Features}
\label{sec-feature}
This section presents the extraction of spectro-temporal deep embedding features and their incorporation into hybrid DNN/TDNN and end-to-end Conformer based ASR systems for auxiliary feature based speaker adaptation.

\begin{figure}[ht]
  \centering
  \includegraphics[scale=0.32]{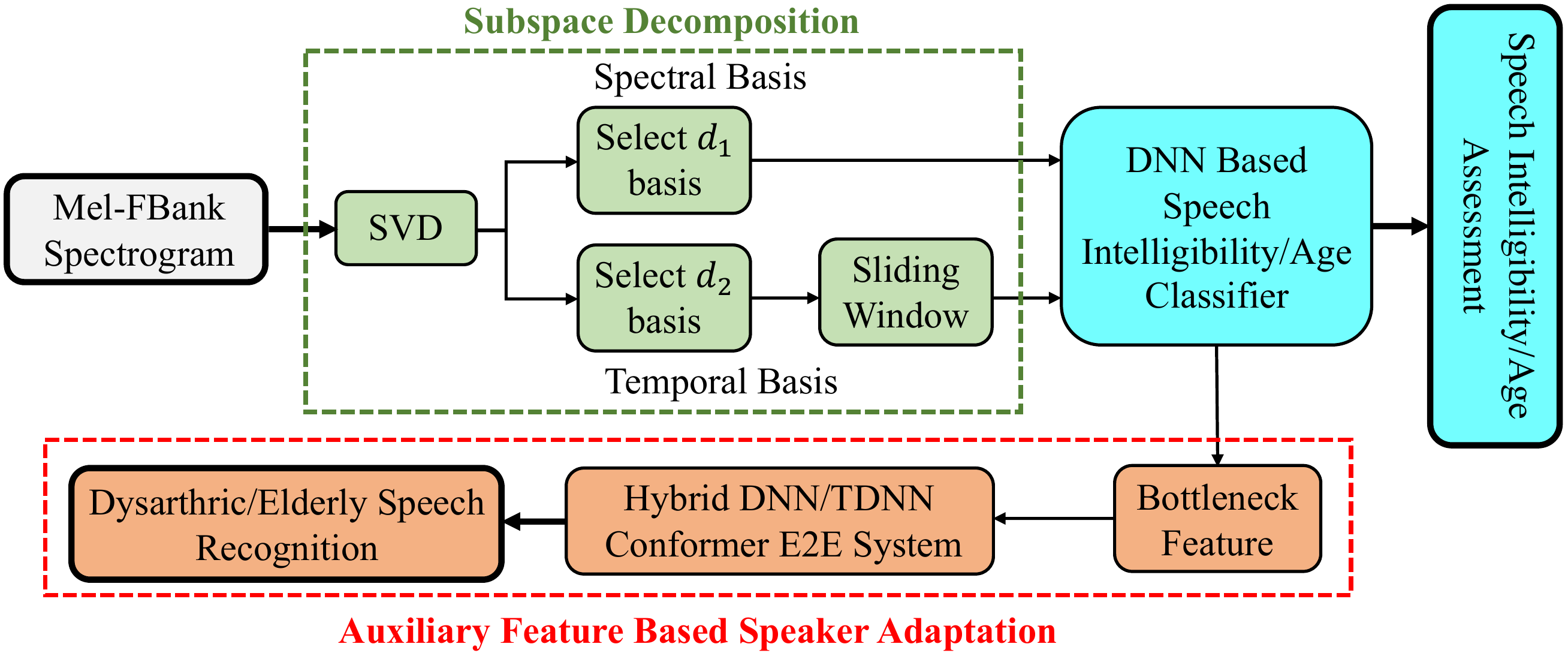}
  \caption{Overall system architecture including from left to right: a) front-end mel-filter bank feature extraction (in grey, top left); b) SVD spectrum decomposition (circled in green, top middle); c) DNN based speech impairment severity or age classification and deep spectro-temporal embedding feature extraction (in light blue, top right); d) auxiliary feature based ASR system adaptation (in orange, bottom).}
  \label{fig:procedure}
\end{figure}

In order to obtain sufficiently discriminative feature representations to capture dysarthric and elderly speaker-level diversity associated with the underlying speech impairment severity level and age information, further supervised learning of deep spectro-temporal embedding features is performed by constructing deep neural network based speech pathology severity or age classifiers. The principal SVD decomposed utterance-level spectral or temporal basis vectors are used as their inputs. More compact, lower dimensional speaker specific spectral and temporal embedding features are then extracted from the resulting impairment severity or age DNN classifiers' bottleneck layers, before being further used as auxiliary embedding features for speaker adaptation of ASR systems. An overall system architecture flow chart covering all the three stages including SVD spectrum decomposition, deep spectral and temporal embedding features extraction and ASR system adaptation using such features is illustrated in Fig.~\ref{fig:procedure}.

\subsection{Extraction of Spectro-Temporal Deep Features}
\label{sec-classifier}

\begin{figure}[ht]
  \centering
  \includegraphics[scale=0.65]{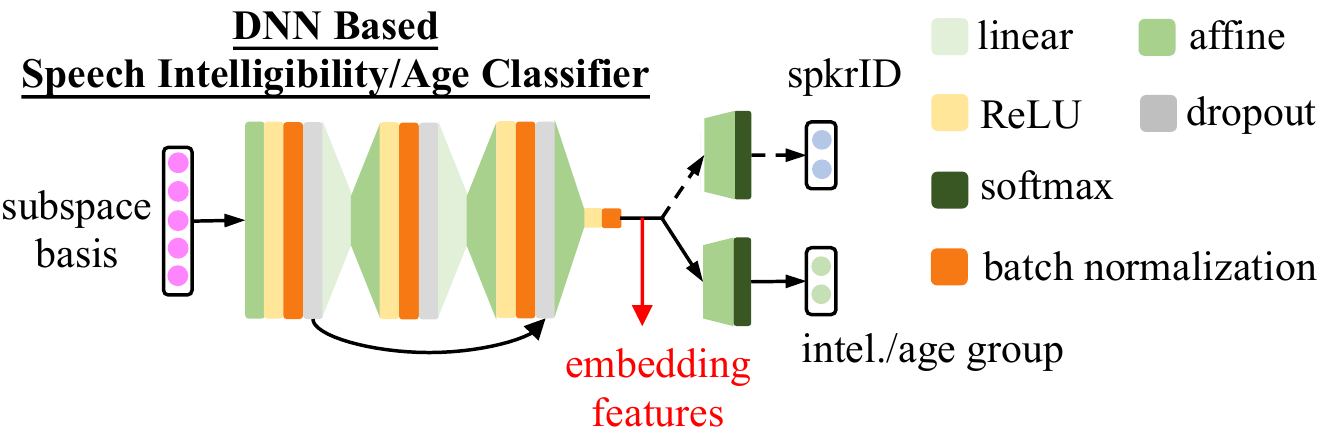}
  \caption{An example DNN based speech intelligibility or age classifier containing a bottleneck layer to extract spectral and temporal embedding features for speaker adaptation.}
  \label{fig:classifier}
\end{figure}

When training the speech impairment severity or age classification DNNs to extract deep spectro-temporoal embedding features, the top-$d$ principal spectral or temporal basis are used as input features to train the respective DNNs sharing the same model architecture shown in Fig.~\ref{fig:classifier}, where either speech pathology severity based on, for example, the speech intelligibility metrics provided by the UASpeech~\cite{kim2008dysarthric} corpus, or the binary aged v.s. non-aged speaker annotation of the DementiaBank Pitt~\cite{becker1994natural} dataset, are used as the output targets. 

The DNN classifier architecture is a fully-connected neural network containing four hidden layers, the first three of which are of $2000$ dimensions, while the last layer contains $25$ dimensions. Each of these hidden layers contains a set of neural operations performed in sequence. These include affine transformation (in green), rectified linear unit (ReLU) activation (in yellow) and batch normalization (in orange), while the outputs of the first layer are connected to those of the third layer via a skip connection. Linear bottleneck projection (in light green) is also applied to the inputs of the middle two hidden layers while dropout operation (in grey) is used on the outputs of the first three hidden layers. Softmax activation (in dark green) is used in the last layer. Further fine-grained speaker-level information can be incorporated into the training cost via a multitask learning (MTL)~\cite{wu2015multi} interpolation between the cross-entropy over speech intelligibility level or age, and that computed over speaker IDs. The outputs of the $25$-dimensional bottleneck (BTN) layer are extracted as compact neural embedding representations of the spectral or temporal basis vectors (bottom right in Fig.~\ref{fig:classifier}).  

When training the DNN speech impairment severity or age classifier using the SVD temporal basis vectors as the input, a frame-level sliding window of $25$ dimensions was applied to the top-$d$ selected temporal basis vectors. Their corresponding $25$-dimensional mean and standard deviation vectors were then computed to serve as the ``average'' temporal basis representations of fixed dimensionality. This within utterance windowed averaging of temporal basis vectors allows dysarthric or elderly speakers who speak of different word contents but exhibit similar patterns of temporal context characteristics such as slower speaking rate and increased pauses to be mapped consistently to the same speech impairment severity or age label. This flexible design is in contrast to conventional speech intelligibility assessment approaches that often require the contents spoken by different speakers to be the same~\cite{berndt1994using,bocklet2013automatic,janbakhshi2020subspace}. It not only facilitates a more practical speech pathology assessment scheme to be applied to unrestricted speech contents of unknown duration, but also the extraction of fixed size temporal embedding features for ASR system adaptation.

\begin{figure*}[ht]
  \centering
    \subfloat[DBANK-iVector]{\includegraphics[width=0.32\textwidth]{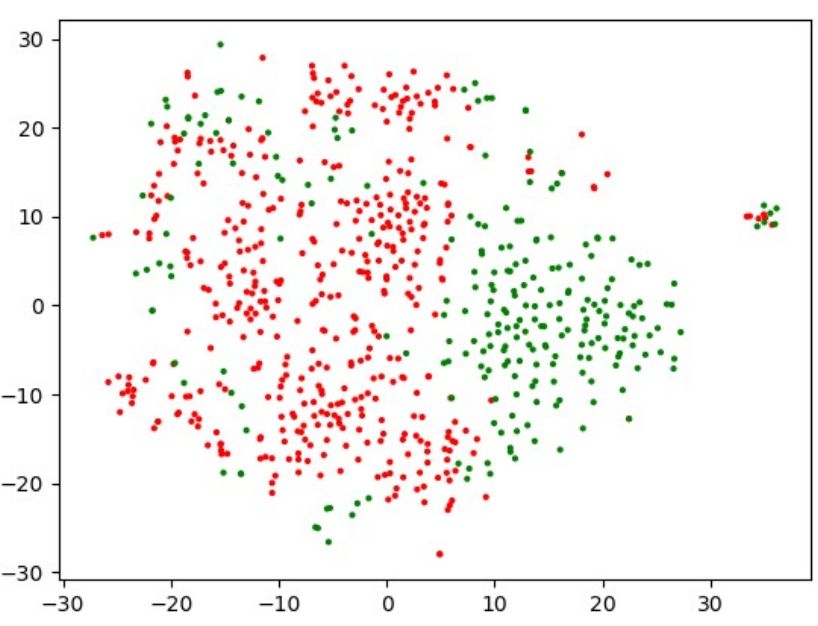}\label{fig:DBANK-iVector}}  
    \hfill
    \subfloat[DBANK-xVector]{\includegraphics[width=0.32\textwidth]{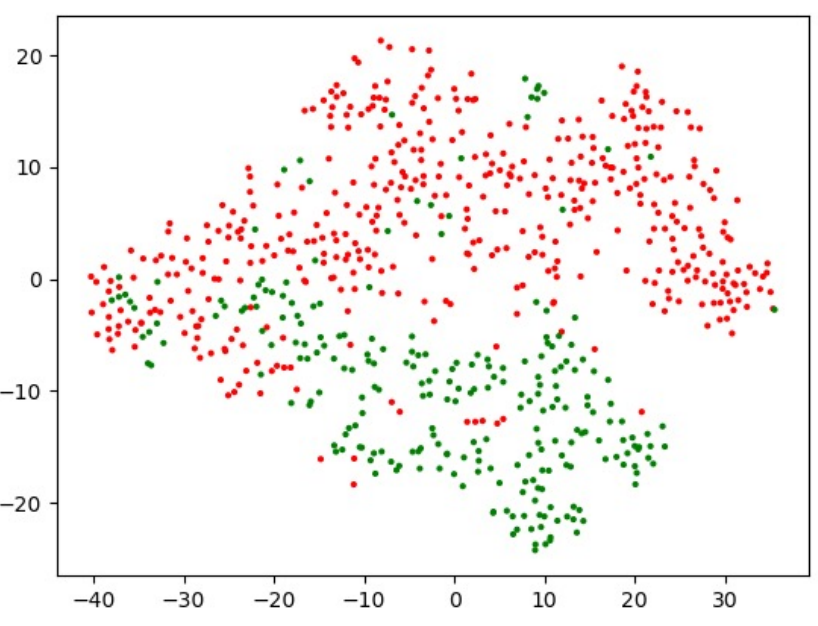}\label{fig:DBANK-xVector}}
    \hfill
    \subfloat[DBANK-BTN Feats.]{\includegraphics[width=0.32\textwidth]{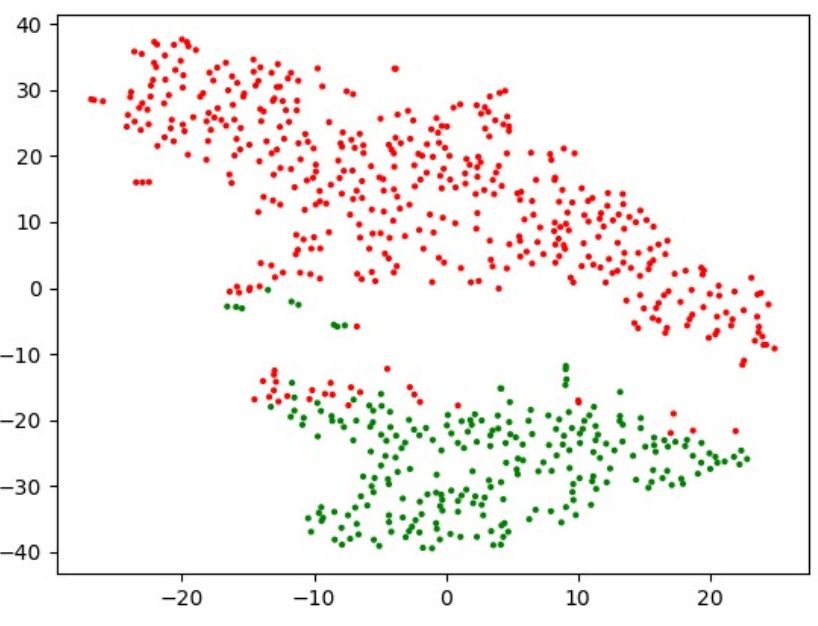}\label{fig:DBANK-spectral}}
    \hfill
    \subfloat[JCMoCA-iVector]{\includegraphics[width=0.32\textwidth]{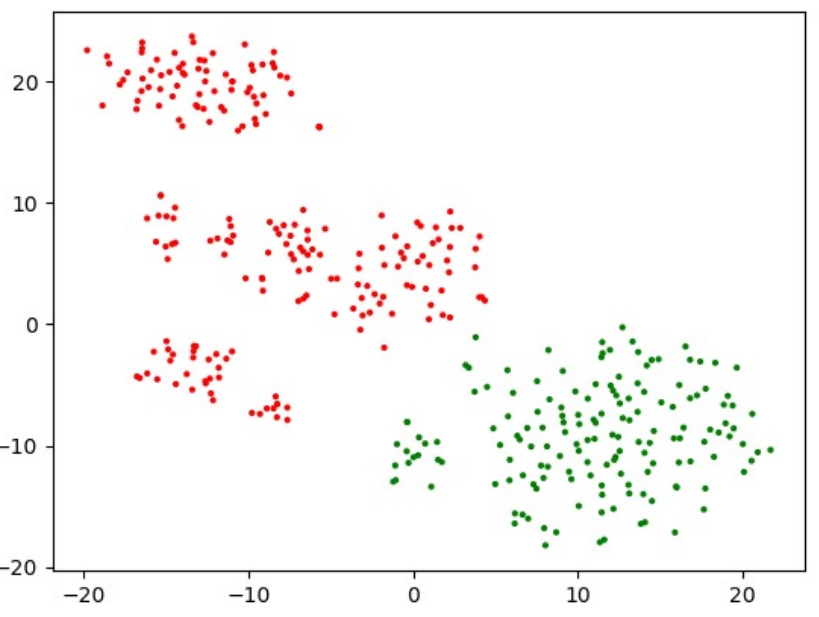}\label{fig:JM-iVector}}  
    \hfill
    \subfloat[JCMoCA-xVector]{\includegraphics[width=0.32\textwidth]{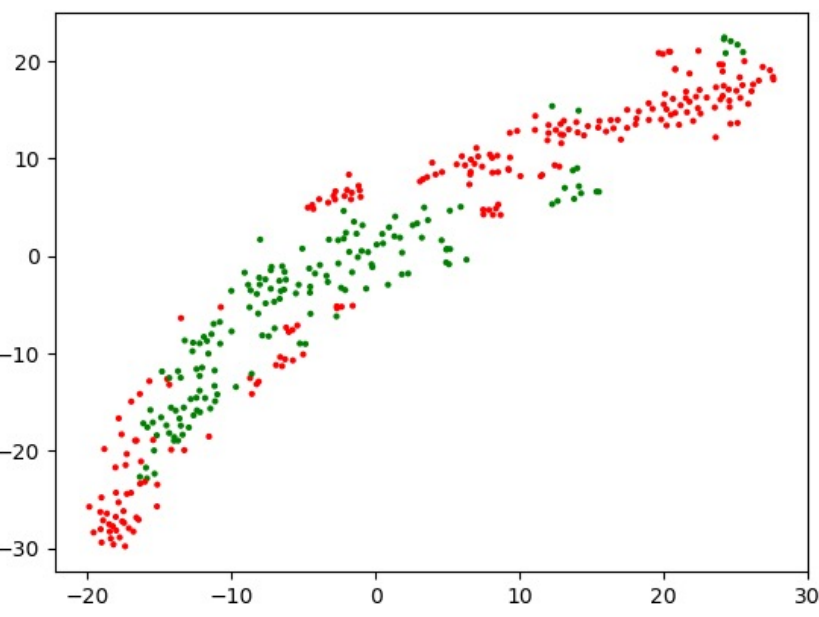}\label{fig:JM-xVector}}
    \hfill
    \subfloat[JCMoCA--BTN Feats.]{\includegraphics[width=0.32\textwidth]{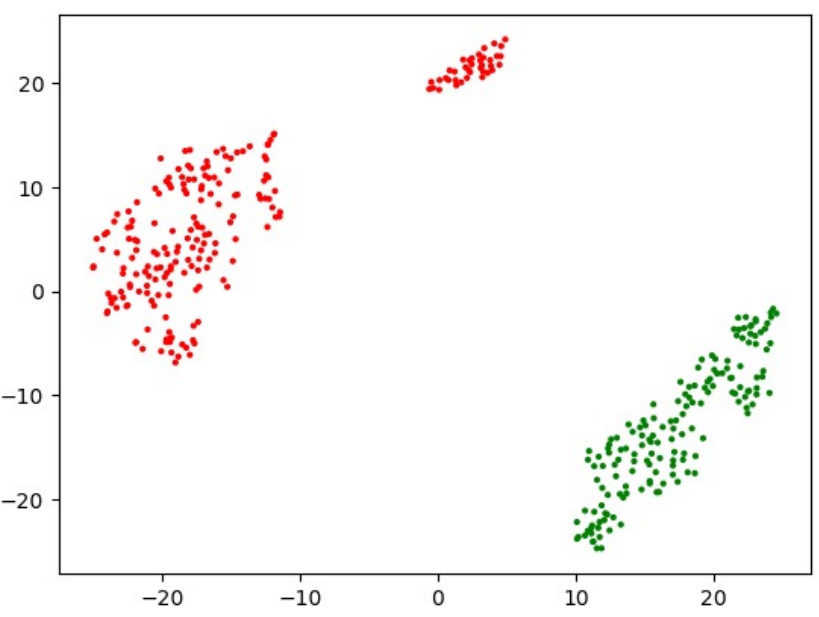}\label{fig:JM-spectral}}
  \caption{T-SNE plot of i-Vectors, x-Vectors and spectral DNN embedding features obtained on the English \textbf{DementiaBank Pitt} (DBANK) corpus with $688$ speakers ($444$ non-aged clinical investigators in red and $244$ aged participants in green) and the Cantonese \textbf{JCCOCC MoCA} (JCMoCA) corpus with $369$ speakers ($211$ non-aged clinical investigators in red and $158$ aged participants in green).}
  \label{t-SNE plot}
\end{figure*}

The speaker-level speech impairment severity or age information can be then captured by the resulting DNN embedding features. For example, visualization using t-distributed stochastic neighbour embedding (t-SNE)~\cite{van2008visualizing} reveals the speaker-level spectral basis neural embedding features averaged over those obtained over all utterances of the same non-aged clinical investigator (in red) or elderly participant (in green) of the DementiaBank Pitt~\cite{becker1994natural} corpus shown in Fig.~\ref{fig:DBANK-spectral} demonstrate much clearer age discrimination than the comparable speaker-level i-Vectors and x-Vectors shown in Fig~\ref{fig:DBANK-iVector} and Fig.~\ref{fig:DBANK-xVector} respectively. Similar trends can also be found on the Cantonese JCCOCC MoCA~\cite{xu2021speaker} corpus designed by a similar data collection protocol based on neuro-physiological interviews comparable to the English DementiaBank Pitt corpus\footnote{Due to the relatively smaller number of speakers included in the UASpeech~\cite{kim2008dysarthric} and TORGO~\cite{rudzicz2012torgo} dysarthric speech corpora (29 and 15 respectively), t-SNE visualization of spectral embedding features are performed on the English DementiaBank Pitt and Cantonese JCCOCC MoCA datasets containing $688$ and $369$ speakers each.}.

\subsection{Use of Spectro-Temporal Deep Features}
\label{sec-use}

\begin{figure*}[ht]
  \centering
    \subfloat[]{\includegraphics[width=0.49\textwidth]{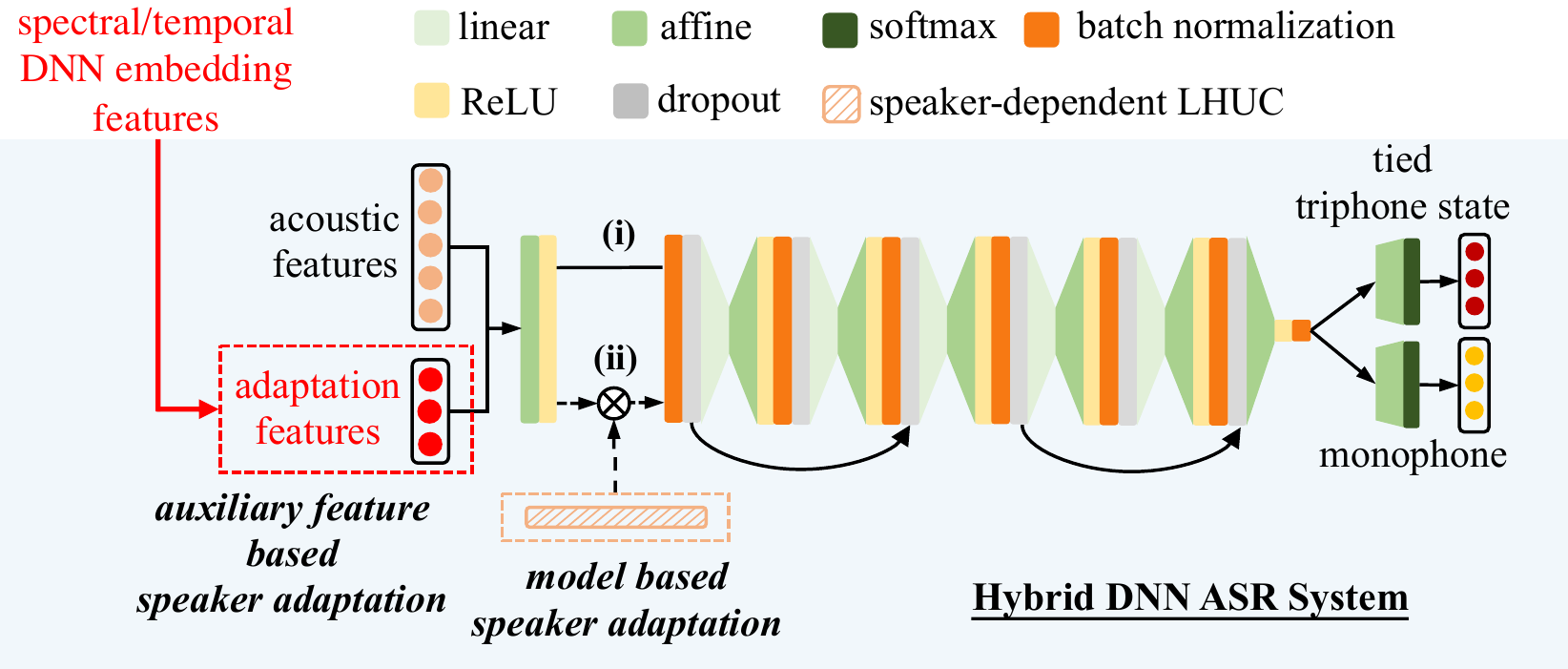}\label{fig:DNN-ASR}}  
  \hfill
  \subfloat[]{\includegraphics[width=0.49\textwidth]{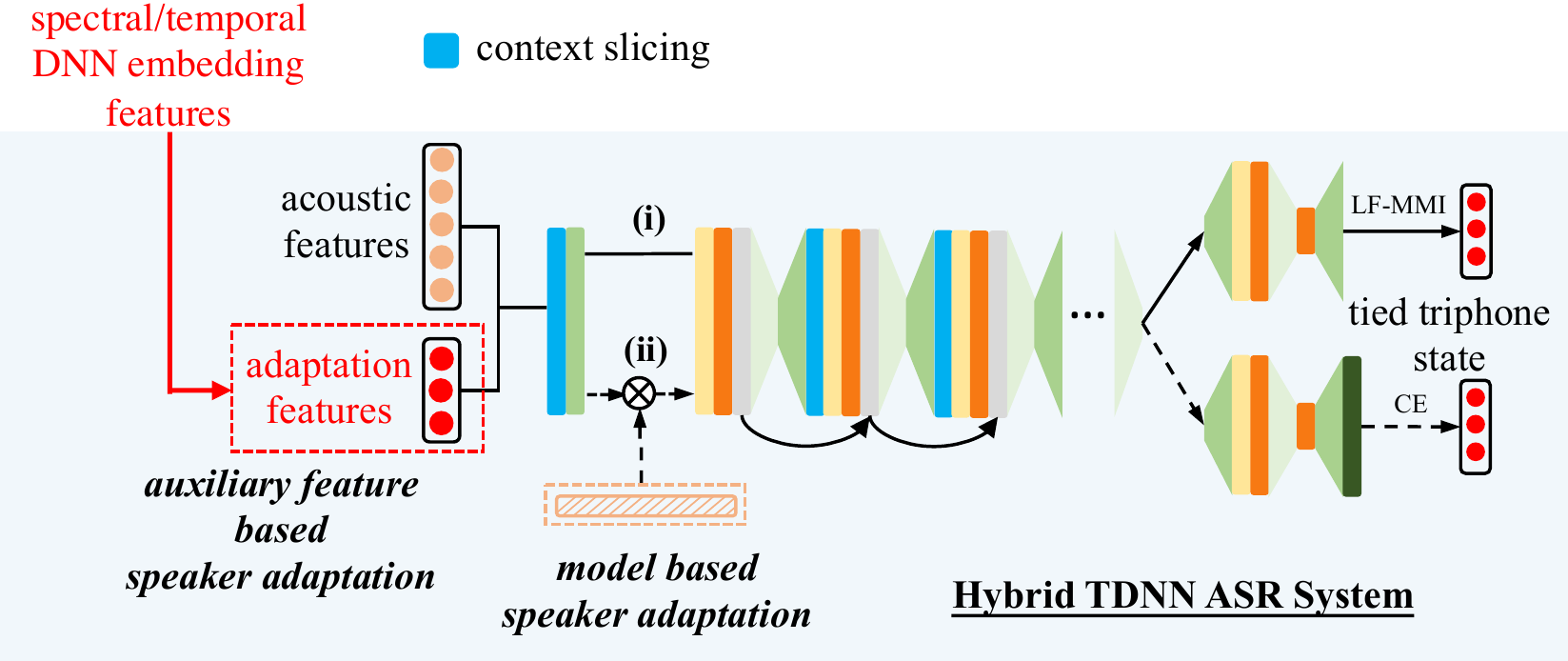}\label{fig:TDNN-ASR}}
  \hfill
\subfloat[]{\includegraphics[width=0.96\textwidth]{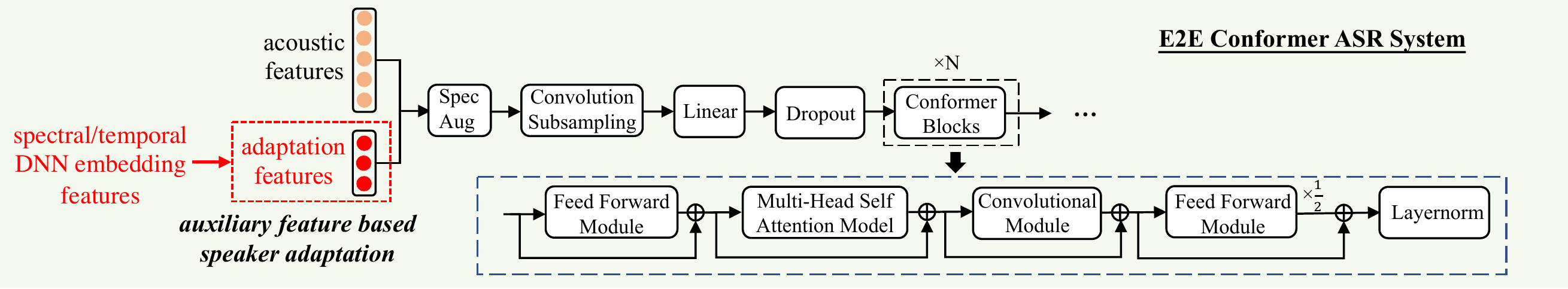}\label{fig:Conformer-ASR}}
  \caption{Incorporation of spectral and temporal deep features at the front end of (a) hybrid DNN~\cite{liu2021recent}, (b) hybrid TDNN~\cite{peddinti2015time} and (c) Conformer~\cite{gulati2020conformer} ASR systems for auxiliary feature based speaker adaptation. For hybrid DNN and TDNN systems, adaption configuration (i) leads to systems using auxiliary feature based adaptation only, while selecting configuration (ii) leads to systems with additional speaker adaptive training using LHUC-SAT~\cite{swietojanski2016learning}.}
  \label{fig:ASR-system}
\end{figure*}

The compact $25$-dimensional spectral and temporal basis embedding features extracted from the DNN speech impairment severity or age classifiers' bottleneck layers presented above in Section~\ref{sec-classifier} are concatenated to the acoustic features at the front-end to facilitate auxiliary feature based speaker adaptation of state-of-the-art ASR systems based on hybrid DNN~\cite{liu2021recent}, hybrid lattice-free maximum mutual information (LF-MMI) trained time delay neural network (TDNN)~\cite{peddinti2015time} or end-to-end (E2E) Conformer models~\cite{gulati2020conformer}, as shown in Fig~\ref{fig:ASR-system}. For hybrid DNN and TDNN systems, model based adaptation using learning unit contributions (LHUC) ~\cite{swietojanski2016learning} can optionally be further applied on top of auxiliary feature based speaker adaptation, as shown in Fig.~\ref{fig:DNN-ASR} and Fig.~\ref{fig:TDNN-ASR} respectively.

\section{Implementation Details}
\label{sec-implementation}
In this section, several key implementation issues associated with the learning and usage of spectro-temporal deep embedding features are discussed. These include the choices of spectro-temporal basis embedding neural network output targets when incorporating speech intelligibility measures or age, the smoothing of the resulting embedding features extracted from such embedding DNNs to ensure the homogeneity over speaker-level characteristics, and the number of principal spectral and temporal basis vectors required for the embedding networks. Ablation studies were conducted on the UASpeech dysarthric speech corpus~\cite{kim2008dysarthric} and the DementiaBank Pitt elderly speech corpus~\cite{becker1994natural}. After speaker independent and speaker dependent speed perturbation based data augmentation~\cite{geng2020investigation,ye2021development}, their respective training data contain approximately $130.1$ hours and $58.9$ hours of speech. After audio segmentation and removal of excessive silence, the UASpeech evaluation data contains $9$ hours of speech while DementiaBank development and evaluation sets of $2.5$ hours and $0.6$ hours of speech respectively were used. Mel-scale filter-bank (FBK) based log amplitude spectra of $40$ channels are used as the inputs of singular value decomposition (SVD) in all experiments of this paper.

\subsection{Choices of Embedding Network Targets}
In the two dysarthric speech corpora, speech pathology assessment measures are provided for each speaker. In the UASpeech data, the speakers are divided into several speech intelligibility subgroups: ``very low'', ``low'', ``mid'' and ``high''~\cite{kim2008dysarthric}. In the TORGO corpus, speech impairment severity measures based on ``severe'', ``moderate'' and ``mild'' are provided~\cite{rudzicz2012torgo}. In the two elderly speech corpora, the role of each speaker during neuro-physiological interview for cognitive impairment assessment is annotated. Each interview is based on a two-speaker conversation involving a non-aged investigator and another aged, elderly participant~\cite{becker1994natural,xu2021speaker}.

By default, the speech intelligibility metrics provided by the UASpeech corpus, or the binary aged v.s. non-aged speaker annotation of the DementiaBank Pitt dataset, are used as the output targets in the following ablation study over embedding target choices. In order to incorporate further speaker-level information, a multitask learning (MTL)~\cite{wu2015multi} style cost function featuring interpolation between the cross-entropy error computed over the speech intelligibility level or age labels, and that computed over speaker IDs can be used. 

As is shown in the results obtained on the UASpeech~\cite{kim2008dysarthric} data in Table~\ref{tab:ablation-UASpeech}, using both the speech intelligibility and speaker ID labels as the embedding targets in multi-task training produced lower word error rates (WERs) across all severity subgroups than using speech intelligibility output targets only (Sys.7 v.s. Sys.6 in Table~\ref{tab:ablation-UASpeech}). The results obtained on the DementiaBank Pitt~\cite{becker1994natural} data in Table~\ref{tab:ablation-DBANK} suggest that there is no additional benefit in adding the speaker information during the embedding process (Sys.7 v.s. Sys.6 in Table~\ref{tab:ablation-DBANK}). Based on these trends, in the main experiments of the following Section~\ref{sec-experiment}, the embedding network output targets exclusively use both speech severity measures and speaker IDs on the UASpeech and TORGO~\cite{rudzicz2012torgo} dysarthric speech datasets, while only binary aged v.s. non-aged labels are used on the DementiaBank Pitt and Cantonese JCCOCC MoCA~\cite{xu2021speaker} elderly speech datasets.

\subsection{Smoothing of Embedding Features}
For auxiliary feature based adaptation techniques including the spectral and temporal basis deep embedding representations considered in this paper, it is vital to ensure the speaker-level homogeneity to be consistently encoded in these features. As both forms of embedding features are computed on individual utterances, additional smoothing is required to ensure such homogeneity, for example, an overall reduction of speech volume of a dysarthric or elderly speaker’s data, to be consistently retained in the resulting speaker embedding representations. To this end, two forms of speaker embedding smoothing are considered in this paper. The first is based on a simple averaging of all utterance-level spectral or temporal embedding features for each speaker. The second smoothing method is based on Latent Dirichlet allocation (LDA)~\cite{blei2003latent} based clustering of utterance-level spectral or temporal embedding features. Following earlier researches~\cite{doulaty2015latent}, a $100$-component Gaussian Mixture Model (GMM) is trained first to quantize the utterance-level spectral or temporal embedding features of the same speaker, before LDA clustering is applied to produce speaker-level features of varying dimensions ranging from $10$, $25$ to $50$.

A general trend observed in the results of both Table~\ref{tab:ablation-UASpeech} and~\ref{tab:ablation-DBANK} is that using spectral embedding feature smoothing, whether via a simple speaker-level averaging (Sys.6 in Table~\ref{tab:ablation-UASpeech} and~\ref{tab:ablation-DBANK}) or LDA clustering (Sys.3-5 in Table~\ref{tab:ablation-UASpeech} and~\ref{tab:ablation-DBANK}), produced better performance than directly using non-smoothed spectral embedding features (Sys.2 in Table~\ref{tab:ablation-UASpeech} and~\ref{tab:ablation-DBANK}). Across both the UASpeech and DementiaBank Pitt tasks, the simpler speaker-level averaging based smoothing (Sys.6 in Table~\ref{tab:ablation-UASpeech} and~\ref{tab:ablation-DBANK}) consistently outperform LDA clustering (Sys.3-5 in Table~\ref{tab:ablation-UASpeech} and~\ref{tab:ablation-DBANK}), and is subsequently used in all experiments of the following Section~\ref{sec-experiment}. 

\begin{table} [ht]
    \caption{Ablation study on the augmented \textbf{UASpeech} corpus~\cite{kim2008dysarthric} with $130.1$h training data. ``SB'', ``TB'' and ``STB'' are in short for spectral or temporal basis vectors and spectral plus temporal basis vectors. ``Seve.'' and ``SpkId'' stand for speech impairment severity group and speaker ID. ``Dim./(d)'' denote the dimensionality and the number of principal spectral or temporal vectors. ``LDA-$10$'', ``LDA-$25$'' and ``LDA-$50$'' denote Latent Dirichlet allocation based clustering features of $10$, $25$ and $50$ dimensions obtained on the embedding features. ``Avg.'' stands for speaker-level averaging of the embedding features. ``O.V." stands for ``overall''.}
    \label{tab:ablation-UASpeech}
    \centering
    \renewcommand\arraystretch{1.0}
    \renewcommand\tabcolsep{2.0pt}
    \scalebox{0.73}{\begin{tabular}{c|c|c|c|c|c|c|cccc|c}
      \hline\hline
          \multirow{3}{*}{Sys.} &
          \multicolumn{4}{c|}{Embed. Network} &
          \multirow{3}{*}{Subspace} & 
          \multirow{3}{*}{Avg.} & 
          \multicolumn{5}{c}{WER} \\
      \cline{2-5}\cline{8-12} 
        & \multicolumn{2}{c|}{Input} & \multicolumn{2}{c|}{Target} & & & \multirow{2}{*}{VL} & \multirow{2}{*}{L} & \multirow{2}{*}{M} & \multirow{2}{*}{H} & \multirow{2}{*}{O.V.} \\
      \cline{2-5}
        & Basis & Dim./(d) & Seve. & SpkId &  &  &  &  &  &  &  \\
      \hline\hline
        1 & \multicolumn{6}{c|}{/} & 66.45  & 28.95  & 20.37 & 9.62  & 28.73 \\ 
      \hline
        2 & \multirow{6}{*}{SB} & \multirow{6}{*}{80/(2)}  & \multirow{6}{*}{\cmark} & \multirow{4}{*}{\xmark} & embed. & \multirow{4}{*}{\xmark} & 66.95 & 32.38 & 21.86 & 11.11 & 30.52 \\ 
        3 & & & & & +LDA-10 &  & 64.22 & 28.10 & 19.13 & 9.07 & 27.62 \\ 
        4 & & & & & +LDA-25 &  & 62.92 & 29.25 & 20.00 & 8.51 & 27.62 \\
        5 & & & & & +LDA-50 &  & 62.62 & 29.22 & 20.03 & 8.42 & 27.52 \\ 
      \cline{1-1}\cline{5-12}
        6 & & & & \xmark & \multirow{2}{*}{embed.} & \multirow{2}{*}{\cmark} & 62.70 & 28.65 & 18.60 & 8.60 & 27.18 \\ 
        7 & & & & \textbf{\cmark} & & & \textbf{61.55} & \textbf{27.52} & \textbf{17.31} & \textbf{8.22} & \textbf{26.26} \\ 
      \hline\hline
      8 & TB & 250/(5) & \multirow{2}{*}{\cmark} & \multirow{2}{*}{\cmark} & \multirow{2}{*}{embed.} & \multirow{2}{*}{\cmark} & 68.52 & 32.24 & 20.98 & 9.46 & 30.09 \\ 
      9 & STB & 330/(2,5) &  &  &  &  & 61.24 & 27.77 & 17.45 & 8.31 & 26.32  \\ 
      \hline\hline
      10 & \multirow{7}{*}{SB} & 40/(1) & \multirow{7}{*}{\cmark} & \multirow{7}{*}{\cmark} & \multirow{7}{*}{embed.} & \multirow{7}{*}{\cmark} & 70.49 & 49.19 & 27.05 & 11.93 & 36.90 \\
      11 &  & 120/(3) &  &  &  &  & 64.50 & 33.01 & 19.47 & 9.82 &  29.27 \\
      12 &  & 160/(4) &  &  &  &  & 72.22 & 47.52 & 20.27 & 9.63 &  34.76 \\
      13 &  & 200/(5) &  &  &  &  & 67.71 & 34.05 & 19.96 & 9.28 &  30.14 \\
      14 &  & 400/(10) &  &  &  &  & 69.82 & 45.98 & 33.23 & 13.83 & 37.76 \\
      15 &  & 800/(20) &  &  &  &  & 74.68 & 45.82 & 28.72 & 11.98 & 37.25 \\
      16 &  & 1600/(40) &  &  &  &  & 71.39 & 44.38 & 29.07 & 11.23 &	36.00 \\
      \hline\hline
    \end{tabular}}
\end{table}

\begin{table} [ht]
    \caption{Ablation study on the augmented \textbf{DementiaBank Pitt} corpus~\cite{becker1994natural} with $58.9$h training data. ``Age'' and ``SpkId'' stand for speaker age group and speaker ID. ``Dev'' and ``Eval'' stand for the development and evaluation sets. ``INV'' and ``PAR'' denote non-aged clinical investigator and aged participant. Other naming conventions follow Table~\ref{tab:ablation-DBANK}.}
    \label{tab:ablation-DBANK}
    \centering
    \renewcommand\arraystretch{1.0}
    \renewcommand\tabcolsep{2.0pt}
    \scalebox{0.73}{\begin{tabular}{c|c|c|c|c|c|c|cc|cc|c}
      \hline\hline
        \multirow{3}{*}{Sys.} &
        \multicolumn{4}{c|}{Embed. Network} &
        \multirow{3}{*}{Subspace} & 
        \multirow{3}{*}{Avg.} & 
        \multicolumn{5}{c}{WER} \\
      \cline{2-5}\cline{8-12} 
        & \multicolumn{2}{c|}{Input} & \multicolumn{2}{c|}{Target} & & & \multicolumn{2}{c|}{Dev} & \multicolumn{2}{c|}{Eval} & \multirow{2}{*}{O.V.} \\
      \cline{2-5}\cline{8-11} 
        & Basis & Dim./(d) & Age & SpkId &  &  & INV & PAR & INV & PAR &  \\
      \hline\hline
        1 & \multicolumn{6}{c|}{/} & 19.91 & 47.93 & 19.76 & 36.66 & 33.80 \\ 
      \hline
       2 & \multirow{6}{*}{SB} & \multirow{6}{*}{160/(4)}  & \multirow{6}{*}{\cmark} &
      \multirow{4}{*}{\xmark} & embed. & \multirow{4}{*}{\xmark} & 19.88 & 45.91 & 17.54 & 33.72 & 32.43 \\ 
       3 & & & & & +LDA-10 &  & 19.31 & 45.50 & 19.31 & 45.50 & 32.25 \\
       4 & & & & & +LDA-25 &  & 19.86 & 45.78 & 19.86 & 45.78 & 32.85 \\
       5 & & & & & +LDA-50 &  & 20.40 & 46.30 & 20.40 & 46.30 & 33.59 \\
      \cline{1-1}\cline{5-12}
       6 & & & & \xmark & \multirow{2}{*}{embed.} & \multirow{2}{*}{\cmark} & 18.61 & \textbf{43.84} & 17.98 & \textbf{33.82} & \textbf{31.12} \\
       7 & & & & \cmark & & & 18.49 & 44.24 & 18.53 & 34.01 & 31.28 \\ 
      \hline\hline
       8 & TB & 160/(4) &\multirow{2}{*}{\cmark} & \multirow{2}{*}{\xmark} & \multirow{2}{*}{embed.} & \multirow{2}{*}{\cmark} & 19.28 & 45.35 & 20.75 & 34.18 & 32.14 \\ 
       9 & STB & 660/(4,10) &  &  &  &  & 20.10 & 46.00 &  20.53 & 35.31 & 32.91 \\ 
      \hline\hline
      10 & \multirow{7}{*}{SB} & 40/(1) & \multirow{7}{*}{\cmark} & \multirow{7}{*}{\xmark} & \multirow{7}{*}{embed.} & \multirow{7}{*}{\cmark} & 18.98 & 44.07 & 19.87 & 33.38 & 31.35 \\
      11 &  & 80/(2) &  &  &  &  & 18.68 & 44.72 & 17.54 & 34.03 & 31.52 \\
      12 &  & 120/(3) &  &  &  &  & 18.36 & 44.39 & 19.64 & 34.71 & 31.44 \\
      13 &  & 200/(4) &  &  &  &  & 18.93 & 44.54 & 18.42 & 33.80 & 31.54 \\
      14 &  & 400/(10) &  &  &  &  & 19.38 & 44.18 & 18.87 & 34.41 & 31.69 \\
      15 &  & 800/(20) &  &  &  &  & 19.90 & 45.47 & 19.31 & 35.04 & 32.54 \\
      16 &  & 1600/(40) &  &  &  &  & 19.82 & 45.55 & 20.53 & 34.12 & 32.42 \\
      \hline\hline
    \end{tabular}}
\end{table}

\subsection{Number of Spectral and Temporal Basis Vectors}
In this part of the ablation study on implementation details, the effect of the number of principal spectral or temporal basis vectors on system complexity and performance is analyzed. Consider selecting the top-$d$ principal SVD spectral and temporal basis components, the input feature dimensionality of the spectral basis embedding (SBE) DNN network is then expressed as $40 \times d$, for example, $80$ dimensions when $d=2$. The temporal basis embedding (TBE) network is $50 \times d$ including both the $25$ dimensional mean and the $25$ dimensional standard deviation vectors both computed over a frame-level sliding window of $25$ dimensions for each of the selected top-$d$ principal temporal basis vector, for example, $250$ dimensions when $d=5$. The input dimensionality of the comparable spectro-temporal basis embedding (STBE) network modelling both forms of bases is then $40 \times d_s + 50 \times d_t$, if further allowing the number of principal spectral components $d_s$ and that of the temporal components $d_t$ to be separately adjusted. 

In the experiments of this section, $d_s$ and $d_t$ are empirically adjusted to be $2$ and $5$ for dysarthric speech (Sys.2-9 in Table~\ref{tab:ablation-UASpeech}) while $4$ and $10$ for elderly speech (Sys.2-9 in Table~\ref{tab:ablation-DBANK}). These settings were found to produce the best adaptation performance when the corresponding set of top principal spectral or temporal basis vectors were used to produce the speaker embedding features. For example, as the results shown in both Table~\ref{tab:ablation-UASpeech} and~\ref{tab:ablation-DBANK} for the UASpeech and DemmentiaBank Pitt datasets, varying the number of principal spectral components from $1$ to $40$ (the corresponding input feature dimensionality ranging from $40$ to $1600$, Sys.10-16 in Table~\ref{tab:ablation-UASpeech} and~\ref{tab:ablation-DBANK}) suggests the optimal number of spectral basis vectors is generally set to be $2$ for the dysarthric speech data (Sys.7 in Table~\ref{tab:ablation-UASpeech}) and $4$ for the elderly speech data (Sys.6 in Table~\ref{tab:ablation-DBANK}) when considering both word error rate (WER) and model complexity.

\section{Experiments}
\label{sec-experiment}

In this experiment section, the performance of our proposed deep spectro-temporal embedding feature based adaptation is investigated on four tasks: the English UASpeech~\cite{kim2008dysarthric} and TORGO~\cite{rudzicz2012torgo} dysarthric speech corpora as well as the English DementiaBank Pitt~\cite{becker1994natural} and Cantonese JCCOCC MoCA~\cite{xu2021speaker} elderly speech datasets. The implementation details discussed in Section~\ref{sec-implementation} are adopted. Data augmentation featuring both speaker independent perturbation of dysarthric or elderly speech and speaker dependent speed perturbation of control healthy or non-aged speech following our previous works~\cite{geng2020investigation,ye2021development} is applied on all of these four tasks. A range of acoustic models that give state-of-the-art performance on these tasks are chosen as the baseline speech recognition systems, including hybrid DNN~\cite{liu2021recent}, hybrid lattice-free maximum mutual information (LF-MMI) trained time delay neural network (TDNN)~\cite{peddinti2015time} and end-to-end (E2E) Conformer~\cite{gulati2020conformer} models. Performance comparison against conventional auxiliary embedding feature based speaker adaptation including i-Vector~\cite{saon2013speaker} and x-Vector~\cite{snyder2018x} is conducted. Model based speaker adaptation using learning hidden unit contributions (LHUC)~\cite{swietojanski2016learning} is further applied on top of auxiliary feature based speaker adaptation. Section~\ref{sec-experiment-dys} presents the experiments on the two dysarthric speech corpora while Section~\ref{sec-experiment-elderly} introduces experiments on the two elderly speech datasets. For all the speech recognition results measured in word error rate (WER) presented in this paper, matched pairs sentence-segment word error (MAPSSWE) based statistical signiﬁcance test~\cite{gillick1989some} was performed at a signiﬁcance level $\alpha=0.05$.

\subsection{Experiments on Dysarthric Speech}
\label{sec-experiment-dys}
\subsubsection{the UASpeech Corpus}
The UASpeech~\cite{kim2008dysarthric} corpus is the largest publicly available and widely used dysarthric speech dataset~\cite{kim2008dysarthric}. It is an isolated word recognition tasks containing approximately $103$ hours of speech recorded from $29$ speakers, among whom $16$ are dysarthric speakers and $13$ are control healthy speakers. It is further divided into 3 blocks Block 1 (B1), Block 2 (B2) and Block 3 (B3) per speaker, each containing the same set of $155$ common words and a different set of $100$ uncommon words. The data from B1 and B3 of all the $29$ speakers are treated as the training set which contains $69.1$ hours of audio and $99195$ utterances in total, and the data from B2 collected of all the $16$ dysarthric speakers (excluding speech from control healthy speakers) are used as the test set containing $22.6$ hours of audio and $26520$ utterances in total. 

After removing excessive silence at both ends of the speech audio segments using a HTK~\cite{young2002htk} trained GMM-HMM system~\cite{yu2018development}, a combined total of $30.6$ hours of audio data from B1 and B3 ($99195$ utterances) were used as the training set, while $9$ hours of speech from B2 ($26520$ utterances) were used for performance evaluation. Data augmentation featuring speed perturbing the dysarthric speech in a speaker independent fashion and the control healthy speech in a dysarthric speaker dependent fashion was further conducted~\cite{geng2020investigation} to produce a $130.1$ hours augmented training set ($399110$ utterances, perturbing both healthy and dysarthric speech). If perturbing dysarthric data only, the resulting augmented training set contains $65.9$ hours of speech ($204765$ utterances).

\subsubsection{the TORGO Corpus}
The TORGO~\cite{rudzicz2012torgo} corpus is a dysarthric speech dataset containing $8$ dysarthric and $7$ control healthy speakers with a totally of approximately $13.5$ hours of audio data ($16394$ utterances). It consists of two parts: $5.8$ hours of short sentence based utterances and $7.7$ hours of single word based utterances. Similar to the setting of the UASpeech corpus, a speaker-level data partitioning was conducted combining all $7$ control healthy speakers' data and two-thirds of the $8$ dysarthric speakers' data into the training set ($11.7$ hours). The remaining one-third of the dysarthric speech was used for evaluation ($1.8$ hours).  After removal of excessive silence, the training and test sets contains $6.5$ hours ($14541$ utterances) and $1$ hour ($1892$ utterances) of speech respectively. After data augmentation with both speaker dependent and speaker independent speed perturbation~\cite{geng2020investigation,hu2022exploit}, the augmented training set contains $34.1$ hours of data ($61813$ utterances). 

\subsubsection{Experiment Setup for the UASpeech Corpus}
Following our previous work~\cite{geng2020investigation,liu2021recent}, the hybrid DNN acoustic models containing six $2000$-dimensional and one $100$-dimensional hidden layers were implemented using an extension to the Kaldi toolkit~\cite{povey2011kaldi}. As is shown in Fig.~\ref{fig:DNN-ASR}, each of its hidden layer contains a set of neural operations performed in sequence. These include affine transformation (in green), rectified linear unit (ReLU) activation (in yellow) and batch normalization (in orange). Linear bottleneck projection (in light green) is applied to the inputs of the five intermediate hidden layers while dropout operation (in grey) is applied on the outputs of the first six hidden layers. Softmax activation (in dark green) is applied in the output layer. Two skip connections feed the outputs of the first hidden layer to those of the third and those of the fourth to the sixth respectively. Multi-task learning (MTL)~\cite{wu2015multi} was used to train the hybrid DNN system with frame-level tied triphone states and monophone alignments obtained from a HTK~\cite{young2002htk} trained GMM-HMM system. The end-to-end (E2E) Conformer systems were implemented using the ESPnet toolkit~\cite{watanabe2018espnet}\footnote{$8$ encoder layers + $4$ decoder layers, feed-forward layer dim = $1024$, attention heads = $4$, dim of attention heads = $256$, interpolated CTC+AED cost.} to directly model grapheme (letter) sequence outputs. $80$-dimensional mel-scale filter-bank (FBK) + $\Delta$ features were used as input for both hybrid DNN and E2E Conformer systems while a $9$-frame context window was used in the hybrid DNN system. The extraction of i-Vector\footnote{Kaldi: egs/wsj/s5/local/nnet3/run\_ivector\_common.sh} and x-Vector\footnote{Kaldi: egs/sre16/v1/local/nnet3/xvector/tuning/run\_xvector\_1a.sh} for UASpeech as well as the three other tasks follow the Kaldi recipe. Following the configurations given in~\cite{christensen2012comparative,yu2018development}, a uniform language model with a word grammar network was used in decoding. Using the spectral basis embedding (SBE) features ($d=2$) and temporal basis embedding (TBE) features ($d=5$) trained on the UASpeech B1 plus B3 data considered here for speaker adaptation, their corresponding dysarthric v.s. control binary utterance-level classification accuracies measured on the B2 data of all $29$ speakers are $99.4$\% and $90.2$\% respectively.

\subsubsection{Experiment Setup for the TORGO Corpus}
The hybrid factored time delay neural network (TDNN) systems containing $7$ context slicing layers were trained following the Kaldi~\cite{povey2011kaldi} chain system setup, as illustrated in Fig.~\ref{fig:TDNN-ASR}. The setup of the E2E graphemic Conformer system was the same as that for UASpeech. $40$-dimensional mel-scale FBK features were used as input for both hybrid TDNN and E2E Conformer systems while a $3$-frame context window was used in the hybrid TDNN system. A $3$-gram language model (LM) trained by all the TORGO transcripts with a vocabulary size of $1.6$k was used during recognition with both the hybrid TDNN and E2E Conformer systems.

\subsubsection{Performance Analysis}

\begin{table}[ht]
    \caption{Performance comparison between the proposed spectral and temporal basis embedding feature based adaptation against i-Vector, x-Vector and LHUC adaptation on the \textbf{UASpeech} test set of $16$ dysarthric speakers. ``$6$M'' and ``$26$M'' refer to the number of model parameters. ``DYS" and ``CTL" in ``Data Aug." column standard for perturbing the dysarthric and the normal speech respectively for data augmentation. ``SBE'' and ``TBE'' denote spectral basis and temporal basis embedding features. ``VL/L/M/H" refer to intelligibility subgroups. $\dag$ denotes a statistically significant improvement ($\alpha=0.05$) is obtained over the comparable baseline i-Vector adapted systems (Sys. 2, 7, 12, 17, 22, 27 and 32).}
    \label{tab:recog-result-UASpeech}
    \centering
    \renewcommand\arraystretch{1.0}
    \renewcommand\tabcolsep{2.0pt}
    \scalebox{0.7}{\begin{tabular}{c|c|c|c|c|c|cccc|c}
        \hline\hline
        \multirow{2}{*}{Sys.} &
        \multirow{2}{*}{\tabincell{c}{Model\\(\# Para.a)}} &
        \multirow{2}{*}{\tabincell{c}{Data\\Aug.}} &
        \multirow{2}{*}{\# Hrs} &
        \multirow{2}{*}{\tabincell{c}{Adapt.\\Feat.}} & 
        \multirow{2}{*}{\tabincell{c}{LHUC\\SAT}} &
        \multicolumn{5}{c}{WER\%} \\
        \cline{7-11} 
         & & & & & & VL & L  & M  & H & O.V. \\
        \hline\hline
        1 & \multirow{10}{*}{\tabincell{c}{Hybrid\\DNN\\(6M)}} & \multirow{10}{*}{\xmark} & \multirow{10}{*}{30.6} & \xmark & \multirow{5}{*}{\xmark} & 69.82 & 32.61 & 24.53 & 10.40 & 31.45 \\
        2 & & & & i-Vector & & 67.25 & 32.70 & 22.56 & 10.11 & 30.46 \\
        3 & & & & x-Vector & & 66.29 & 30.00 & 22.23 & 9.29 &  29.40 \\
        4 & & & & SBE  & & \textbf{64.43$^\dag$} & \textbf{29.71$^\dag$} & \textbf{19.84$^\dag$} & \textbf{8.57$^\dag$} & \textbf{28.05$^\dag$} \\
        5 & & & & SBE+TBE & & \textbf{64.54$^\dag$} & \textbf{29.13$^\dag$} & \textbf{18.90$^\dag$} & \textbf{8.69$^\dag$} & \textbf{27.83$^\dag$} \\
        \cline{1-1}\cline{5-11}
        6 & & & & \xmark & \multirow{5}{*}{\cmark} & 64.39 & 29.88 & 20.27 & 8.95 & 28.29 \\
        7 & & & & i-Vector & & 64.31 &  29.46 & 18.39 & 8.70 & 27.72 \\
        8 & & & & x-Vector & & 64.95 &  29.28 & 19.56 & 8.58 & 27.99 \\
        9 & & & & SBE  & & 63.40 & \textbf{28.90$^\dag$} & 18.64 & \textbf{8.13$^\dag$} & \textbf{27.24$^\dag$}  \\
        10 & & & & SBE+TBE & & \textbf{63.01$^\dag$} & \textbf{28.10$^\dag$} & 18.25 & \textbf{8.22$^\dag$} & \textbf{26.90$^\dag$} \\
        \hline\hline
        11 & \multirow{10}{*}{\tabincell{c}{Hybrid\\DNN\\(6M)}} & \multirow{10}{*}{DYS} & \multirow{10}{*}{65.9} & \xmark & \multirow{5}{*}{\xmark} & 68.43 & 29.60 & 21.37 & 10.44 & 29.79 \\
        12 & & & & i-Vector  & & 66.06 & 31.16 & 20.27 & 8.86 & 28.95 \\
        13 & & & & x-Vector  & & 64.22 & 29.00 & 21.27 & 9.23 & 28.53 \\
        14 & & & & SBE  & & \textbf{62.56$^\dag$} & \textbf{28.33$^\dag$} & \textbf{18.21$^\dag$} & 9.01 & \textbf{27.13$^\dag$} \\
        15 & & & & SBE+TBE & & \textbf{61.40$^\dag$} & \textbf{28.93$^\dag$} & \textbf{18.74$^\dag$} & 9.00 & \textbf{27.14$^\dag$} \\
        \cline{1-1}\cline{5-11}
        16 & & & & \xmark & \multirow{5}{*}{\cmark} & 60.99 & 28.20 & 18.86 & 8.41 & 26.69 \\
        17 & & & & i-Vector & & 62.15 & 28.78 &	18.52 &	8.30 & 26.98 \\
        18 & & & & x-Vector & & 63.58 & 28.91 &	20.33 &	8.29 & 27.66\\
        19 & & & & SBE  & & 60.98 & \textbf{27.29$^\dag$} & \textbf{17.96$^\dag$} & 8.54 & \textbf{26.32$^\dag$} \\
        20 & & & & SBE+TBE & & \textbf{60.23$^\dag$} & \textbf{27.87$^\dag$} & 18.27 & \textbf{8.67$^\dag$} & \textbf{26.41$^\dag$} \\
        \hline\hline
        21 & \multirow{10}{*}{\tabincell{c}{Hybrid\\DNN\\(6M)}} & \multirow{10}{*}{\tabincell{c}{DYS\\+\\CTL}} & \multirow{10}{*}{130.1} & \xmark & \multirow{5}{*}{\xmark} & 66.45 & 28.95 & 20.37 & 9.62 & 28.73  \\
        22 & & & & i-Vector & & 65.52 & 30.63 & 19.27 & 8.60 & 28.42 \\
        23 & & & & x-Vector & & 64.50 & 28.00 & 19.94 & 8.48 & 27.82 \\
        24 & & & & SBE  & & \textbf{61.55$^\dag$} & \textbf{27.52$^\dag$} & \textbf{17.31$^\dag$} & \textbf{8.22$^\dag$} & \textbf{26.26$^\dag$} \\
        25 & & & & SBE+TBE & & \textbf{61.24$^\dag$} & \textbf{27.77$^\dag$} & \textbf{17.45$^\dag$} & 8.31 & \textbf{26.32$^\dag$} \\
        \cline{1-1}\cline{5-11}
        26 & & & & \xmark & \multirow{5}{*}{\cmark} & 62.50 & 27.26 & 18.41 & 8.04 & 26.55 \\
        27 & & & & i-Vector & & 61.60 &  28.61 & 17.94 & 8.06 &  26.63 \\
        28 & & & & x-Vector & & 62.83 & 28.84 & 18.09 & 7.93 &  26.93 \\
        29 & & & & SBE & & \textbf{59.30$^\dag$} & \textbf{26.25$^\dag$} & \textbf{16.25$^\dag$} & \textbf{7.60$^\dag$} & \textbf{25.05$^\dag$} \\
        30 & & & & SBE+TBE & & 60.92 & \textbf{27.11$^\dag$} & \textbf{17.00$^\dag$} & \textbf{7.52$^\dag$} & \textbf{25.73$^\dag$} \\
        \hline\hline
        31 & \multirow{5}{*}{\tabincell{c}{Conformer\\(19M)}} & \multirow{5}{*}{\tabincell{c}{DYS\\+\\CTL}}  & \multirow{5}{*}{130.1} & \xmark & \multirow{5}{*}{\xmark} & 66.77 & 49.39 & 46.47 & 42.02 & 50.03 \\
        32 & & & & i-Vector & & 69.32 & 51.05 & 47.23 & 41.81 & 51.07 \\
        33 & & & & x-Vector & & 70.57 & 52.78 & 48.72 & 42.42 & 52.27 \\
        34 & & & & SBE  & & \textbf{65.95$^\dag$} & \textbf{47.79$^\dag$} & \textbf{44.94$^\dag$} & \textbf{41.32$^\dag$} & \textbf{48.91$^\dag$} \\
        35 & & & & SBE+TBE & & \textbf{65.57$^\dag$} &  \textbf{48.13$^\dag$} & \textbf{45.49$^\dag$} & \textbf{41.45$^\dag$} & \textbf{49.06$^\dag$} \\
        \hline\hline
    \end{tabular}}
\end{table}

The performance of the proposed spectro and temporal deep feature based adaptation is compared with that obtained using conventional i-Vector~\cite{saon2013speaker} and x-Vector~\cite{snyder2018x} based adaptation, as shown in Table~\ref{tab:recog-result-UASpeech}. Sys.1-30 were trained using the hybrid DNN system~\cite{liu2021recent}, where Sys.1-10 were trained on the $30.6$h non-augmented training set, Sys.11-20 on the $65.9$h training set augmented by speed perturbing the dysarthric speech only, and Sys.21-30 on the $130.1$h training set augmented by speed perturbing both the dysarthric and control healthy speech~\cite{geng2020investigation}. Sys.31-35 were trained using the E2E Conformer system on the $130.1$h augmented training set. The following trends can be observed: 

i) The proposed spectral and temporal deep feature adapted systems consistently outperformed the comparable baseline speaker independent (SI) systems across all speech intelligibility subgroups with different amount of training data and baseline ASR system settings (Sys.4-5 v.s. Sys.1, Sys.14-15 v.s. Sys.11, Sys.24-25 v.s. Sys.21 and Sys.34-35 v.s. Sys.31) by up to $3.62\%$ absolute ($11.51\%$ relative) statistically significant reduction in overall WER (Sys.5 v.s. Sys.1). 

ii) When compared with conventional i-Vector and x-Vector based adaptation, our proposed spectro-temporal deep feature adapted systems consistently produced lower WERs across very low, low and mid speech intelligibility subgroups (Sys.4-5 v.s. Sys.2-3, Sys.14-15 v.s. Sys.12-13, Sys.24-25 v.s. Sys.22-23 and Sys.34-35 v.s. Sys.32-33). A statistically significant overall WER reduction of up to $2.63\%$ absolute ($8.63\%$ relative) was obtained (Sys.5 v.s. Sys.2). 

iii) When further combined with model based speaker adaptation via LHUC, the spectro-temporal deep feature adapted systems consistently achieved better performance than the systems with no auxiliary feature based adaptation or i-Vector and x-Vector based adaptation (Sys.9-10 v.s. Sys.6-8, Sys.19-20 v.s. Sys.16-18 and Sys.29- 30 v.s. Sys.26-28). A statistically significant  reduction in overall WER by up to $1.58\%$ absolute ($5.93\%$ relative) was obtained (Sys.29 v.s. Sys.27). 

iv) Compared with using spectral basis embedding features (SBE) only in adaptation, using both spectral and temporal embedding features (SBE+TBE) leads to comparable performance but no consistent benefit. For example, marginal performance improvements were obtained on the non-augmented $30.6$h training set (Sys.5 v.s. Sys.4 and Sys.10 v.s. Sys.9) while small performance degradation found on the other augmented training sets for both hybrid DNN (Sys.15 v.s. Sys.14, Sys.20 v.s. Sys.19, Sys.25 v.s. Sys.24 and Sys.30 v.s. Sys.29) and E2E Conformer systems (Sys.35 v.s. Sys.34). Based on these observations, only spectral basis embedding feature (SBE) based adaptation are considered in the remaining experiments of this paper. 

A comparison between previously published systems on the UASpeech corpus and our system is shown in Table~\ref{tab:compare}. To the best of our knowledge, this is the lowest WER obtained by ASR systems published so far on the UASpeech test set of 16 dysarthric speakers in the literature. 

\begin{table}[ht]
  \caption{A comparison between published systems on \textbf{UASpeech} and our system. Here ``DA'' refers to data augmentation and ``GAN'' stands for generative adversarial network. }
  \label{tab:compare}
  \centering
  \renewcommand\arraystretch{1}
  \scalebox{0.8}{\begin{tabular}{cc}
  \toprule
    Systems   &  WER\%   \\
  \midrule
    Sheffield-2013 Cross domain augmentation \cite{christensen2013combining}    & 37.50 \\
    Sheffield-2015 Speaker adaptive training  \cite{sehgal2015model}    & 34.80  \\
    CUHK-2018 DNN System Combination \cite{yu2018development}         & 30.60  \\
    Sheffield-2020 Fine-tuning CNN-TDNN speaker adaptation \cite{xiong2020source} & 30.76 \\
    CUHK-2020 DNN + DA + LHUC SAT \cite{geng2020investigation} & 26.37 \\ 
    CUHK-2021 LAS + CTC + Meta-learning + SAT \cite{wang2021improved} & 35.00 \\
    CUHK-2021 QuartzNet + CTC + Meta-learning + SAT \cite{wang2021improved} & 30.50 \\
    CUHK-2021 DNN + GAN DA \cite{jin2021adversarial} & 25.89 \\
    CUHK-2021 NAS DNN + DA + LHUC SAT + AV fusion \cite{liu2021recent} & 25.21 \\
    \textbf{DA + SBE Adapt + LHUC SAT (Table~\ref{tab:recog-result-UASpeech}, Sys.29)}  & \textbf{25.05} \\ 
   \bottomrule
    \end{tabular}}
\end{table}

On a comparable set of experiments conducted on the on the TORGO~\cite{rudzicz2012torgo} corpus using the $34.1$h augmented training set shown in Table~\ref{tab:recog-result-TORGO}, trends similar to those found the on UASpeech task in Table~\ref{tab:recog-result-UASpeech} can be observed. Compared with the i-Vector adapted systems, a statistically significant overall WER reduction by up to $1.27\%$ absolute ($14\%$ relative) (Sys.4 v.s. Sys.2) can be obtained using the spectral embedding feature (SBE) adapted TDNN systems. The SBE adapted Conformer system outperformed its i-Vector baseline statistically significantly by $1.98\%$ absolute ($14.01\%$ relative) (Sys.12 v.s. Sys.10). 

\begin{table}[ht]
    \caption{Performance comparison between the proposed spectral basis embedding feature based adaptation against i-Vector, x-Vector and LHUC adaptation on the \textbf{TORGO} test set of $8$ dysarthric speakers. ``$10$M'' and ``$18$M'' refer to the number of model parameters. ``DYS + CTL" in ``Data Aug." column denotes perturbing both dysarthric and normal speech in data augmentation. ``SBE'' denote spectral basis embedding features. ``Seve./Mod./Mild" refer to the speech impairment severity levels: severe, moderate and mild. $\dag$ denotes a statistically significant improvement ($\alpha=0.05$) is obtained over the comparable baseline i-Vector adapted systems (Sys. 2, 6 and 10).}
    \label{tab:recog-result-TORGO}
    \centering
    \renewcommand\arraystretch{1.0}
    \renewcommand\tabcolsep{2.0pt}
    \scalebox{0.83}{\begin{tabular}{c|c|c|c|c|c|ccc|c}
        \hline\hline
        \multirow{2}{*}{Sys.} &
        \multirow{2}{*}{\tabincell{c}{Model\\(\# Para.)}} &
        \multirow{2}{*}{\tabincell{c}{Data\\Aug.}} &
        \multirow{2}{*}{\# Hrs} &
        \multirow{2}{*}{\tabincell{c}{Adapt.\\Feat.}} & 
        \multirow{2}{*}{\tabincell{c}{LHUC}} &
        \multicolumn{4}{c}{WER\%} \\
        \cline{7-10} 
         & & & & & & Seve. & Mod. & Mild & O.V. \\
        \hline\hline
        1 & \multirow{8}{*}{\tabincell{c}{Hybrid\\TDNN\\(10M)}} & \multirow{8}{*}{\tabincell{c}{DYS\\+\\CTL}} & \multirow{8}{*}{34.1} & \xmark & \multirow{4}{*}{\xmark} & 12.80 &  8.78 &  3.64 & 9.47 \\
        2 & & & & i-Vector & & 13.82 &  5.92 &  2.40 & 9.07 \\
        3 & & & & x-Vector & & 12.76 &  5.31 &  3.17 & 8.60 \\
        4 & & & & SBE  & & \textbf{11.67$^\dag$}& \textbf{4.59$^\dag$} &  2.86 &  \textbf{7.80$^\dag$} \\
        \cline{1-1}\cline{5-10}
        5 & & & & \xmark & \multirow{4}{*}{\cmark} & 12.60 &    8.78 &  3.64 & 9.36 \\
        6 & & & & i-Vector & & 13.74 & 5.82 & 2.55 & 9.04 \\
        7 & & & & x-Vector & & 12.68 & 5.20 & 3.02 & 8.50 \\
        8 & & & & SBE & & \textbf{11.71$^\dag$}  & \textbf{4.29$^\dag$}  & 2.86 & \textbf{7.76$^\dag$}  \\
        \hline\hline
        9 & \multirow{4}{*}{\tabincell{c}{Conformer\\(18M)}} & \multirow{4}{*}{\tabincell{c}{DYS\\+\\CTL}}  & \multirow{4}{*}{34.1} & \xmark & \multirow{4}{*}{\xmark} & 21.66 & 6.22 & 4.10 & 13.67 \\
        10 & & & & i-Vector & & 22.15 & 7.44 &  3.94 & 14.13 \\
        11 & & & & x-Vector & & 20.52 & 7.55 &  3.71 & 13.25 \\
        12 & & & & SBE & & \textbf{18.69$^\dag$}  & 6.83 &  3.71 &  \textbf{12.15$^\dag$}  \\
        \hline\hline
    \end{tabular}}
\end{table}

\subsection{Experiments on Elderly Speech}
\label{sec-experiment-elderly}
\subsubsection{the DementiaBank Pitt Corpus}
The DementiaBank Pitt~\cite{becker1994natural} corpus contains approximately $33$ hours of audio data recorded over interviews between the $292$ elderly participants and the clinical investigators. It is further split into a $27.2$h training set, a $4.8$h development and a $1.1$h evaluation set for ASR system development. The evaluation set is based on exactly the same $48$ speakers' Cookie (picture description) task recordings as those in the ADReSS\footnote{http://www.homepages.ed.ac.uk/sluzﬁl/ADReSS/}~\cite{luz2020alzheimer} test set, while the development set contains the remaining recordings of these speakers in other tasks if available. The training set contains $688$ speakers ($244$ elderly participants and $444$ investigators), while the development set includes $119$ speakers ($43$ elderly participants and $76$ investigators) and the evaluation set contains $95$ speakers ($48$ elderly participants and $47$ investigators). After removal of excessive silence~\cite{ye2021development}, the training set contains $15.7$ hours of audio data ($29682$ utterances) while the development and evaluation sets contain $2.5$ hours ($5103$ utterances) and $0.6$ hours ($928$ utterances) of audio data respectively. Data augmentation featuring speaker independent speed perturbation of elderly speech and elderly speaker dependent speed perturbation of non-aged investigators' speech~\cite{ye2021development} produced an $58.9$h augmented training set ($112830$ utterances).

\subsubsection{the JCCOCC MoCA Corpus}
The Cantonese JCCOCC MoCA corpus contains conversations recorded from cognitive impairment assessment interviews between 256 elderly participants and the clinical investigators \cite{xu2021speaker}. The training set contains $369$ speakers ($158$ elderly participants and $211$ investigators) with a duration of $32.4$ hours. The development and evaluation sets each contains speech recorded from $49$ elderly speakers. After removal of excessive silence, the training set contains $32.1$ hours of speech ($95448$ utterances) while the development and evaluation sets contain $3.5$ hours ($13675$ utterances) and $3.4$ hours ($13414$ utterances) of speech respectively. After data augmentation following approaches similar to those adopted on the DementiaBank Pitt corpus~\cite{ye2021development}, the augmented training set consists of $156.9$ hours of speech ($389409$ utterances).

\subsubsection{Experiment Setup for the DementiaBank Pitt Corpus}
Following the Kaldi~\cite{povey2011kaldi} chain system setup, the hybrid TDNN system shown in Fig.~\ref{fig:TDNN-ASR} contain $14$ context slicing layers with a $3$-frame context. $40$-dimensional mel-scale FBK features were used as input for all systems. For both the hybrid TDNN and E2E graphemic Conformer systems\footnote{$12$ encoder layers + $12$ decoder layers, feed-forward layer dim = $2048$, attention heads = $4$, dim of attention heads = $256$, interpolated CTC+AED cost.}, a word level $4$-gram LM was trained following the settings of our previous work~\cite{ye2021development} and a $3.8$k word recognition vocabulary covering all the words in the DementiaBank Pitt corpus was used in recognition. Using the spectral basis embedding (SBE) features ($d = 4$) considered here for speaker adaptation, the corresponding aged v.s. non-aged (participant v.s. investigator) utterance-level classification accuracy on the combined development plus evaluation set is $84.9\%$. 

\subsubsection{Experiment Setup for the JCCOCC MoCA Corpus}
The architecture of the hybrid TDNN and E2E graphemic (character) Conformer systems were the same as those for the DementiaBank Pitt corpus above. $40$-dimensional mel-scale FBK features were used as input for all systems. A word level $4$-gram language model with Kneser-Ney smoothing was trained on the transcription of the JCCOCC MoCA corpus ($610$k words) using the SRILM toolkit~\cite{stolcke2002srilm} and a $5.2$k recognition vocabulary covering all the words in the JCCOCC MoCA corpus was used.

\subsubsection{Performance Analysis}
The performance comparison between the proposed spectral deep feature based adaptation against traditional i-Vector~\cite{saon2013speaker} and x-Vector~\cite{snyder2018x} based adaptation using either hybrid TDNN~\cite{peddinti2015time} or E2E Conformer~\cite{gulati2020conformer} systems on the DimentiaBank Pitt corpus with the $58.9$h augmented training set is shown in Table~\ref{tab:recog-result-DBANK}, where trends similar to those found on the the dysarthric speech experiments in Table~\ref{tab:recog-result-UASpeech} and~\ref{tab:recog-result-TORGO} can be observed:

i) The proposed spectral basis embedding feature (SBE) adapted systems consistently outperform the comparable baseline speaker independent (SI) systems with or without model based speaker adaptation using LHUC (Sys.4 v.s. Sys.1, Sys.8 v.s. Sys.5 and Sys.12 v.s. Sys.9) by up to $3.17\%$ absolute ($9.81\%$ relative) overall WER reduction (Sys.8 v.s. Sys.5). 

ii) When compared with conventional i-Vector and x-Vector based adaptation, our proposed SBE feature adapted systems consistently produced lower WERs with or without model based speaker adaptation using LHUC (Sys.4 v.s. Sys.2-3, Sys.8 v.s. Sys.6-7 and Sys.12 v.s. Sys.10-11). A statistically significant overall WER reduction of $2.57\%$ absolute ($8.1\%$relative) was obtained (Sys.8 v.s. Sys.6).

\begin{table}[ht]
    \caption{Performance comparison between the proposed spectral basis embedding feature based adaptation against i-Vector, x-Vector and LHUC adaptation on the \textbf{DementiaBank Pitt} corpus. ``$18$M'' and ``$52$M'' refer to the number of model parameters. ``SBE'' denote spectral basis embedding features. ``Dev'' and ``Eval'' stand for the development and evaluation sets. ``INV'' and ``PAR'' refer to clinical investigator and elderly participant. $\dag$ denotes a statistically significant improvement ($\alpha=0.05$) is obtained over the comparable baseline i-Vector adapted systems (Sys. 2, 6 and 10).}
    \label{tab:recog-result-DBANK}
    \centering
    \renewcommand\arraystretch{1.0}
    \renewcommand\tabcolsep{2.0pt}
    \scalebox{0.73}{\begin{tabular}{c|c|c|c|c|c|cc|cc|c}
        \hline\hline
        \multirow{3}{*}{Sys.} &
        \multirow{3}{*}{\tabincell{c}{Model\\(\# Para.)}} &
        \multirow{3}{*}{\tabincell{c}{Data\\Aug.}} &
        \multirow{3}{*}{\# Hrs} &
        \multirow{3}{*}{\tabincell{c}{Adapt.\\Feat.}} & 
        \multirow{3}{*}{\tabincell{c}{LHUC\\SAT}} &
        \multicolumn{4}{c}{WER\%} \\
        \cline{7-11} 
         & & & & & & \multicolumn{2}{c|}{Dev} & \multicolumn{2}{c|}{Eval} & \multirow{2}{*}{O.V.} \\ 
        \cline{7-10} 
         & & & & & & INV & PAR & INV & PAR & \\  
        \hline\hline
        1 & \multirow{8}{*}{\tabincell{c}{Hybrid\\TDNN\\(18M)}} & \multirow{8}{*}{\cmark} & \multirow{8}{*}{58.9} & \xmark & \multirow{4}{*}{\xmark} & 19.91 & 47.93 & 19.76 & 36.66 & 33.80 \\
        2 & & & & i-Vector & & 19.97 & 46.76 & 18.20 & 37.01 & 33.37 \\
        3 & & & & x-Vector & & 18.01 & 46.42 & 18.76 & 37.62 & 32.56 \\
        4 & & & & SBE  & & 18.61$^\dag$ & \textbf{43.84$^\dag$} & 17.98 & \textbf{33.82$^\dag$} & \textbf{31.12$^\dag$} \\
        \cline{1-1}\cline{5-11}
        5 & & & & \xmark & \multirow{4}{*}{\cmark} & 19.26 & 45.49 & 18.42 & 35.44 & 32.33 \\
        6 & & & & i-Vector & & 18.62 & 44.70 & 17.98 & 35.38 & 31.73 \\
        7 & & & & x-Vector & & 17.93 & 45.76 & 16.76 & 36.11 & 31.95 \\
        8 & & & & SBE & & 17.41$^\dag$ & \textbf{40.94$^\dag$} & 17.98 & \textbf{31.89$^\dag$} & \textbf{29.16$^\dag$} \\
        \hline\hline
        9 & \multirow{4}{*}{\tabincell{c}{Conformer\\(52M)}} & \multirow{4}{*}{\tabincell{c}{\cmark}}  & \multirow{4}{*}{58.9} & \xmark & \multirow{4}{*}{\xmark} & 20.97 & 48.71 & 19.42 & 36.93 & 34.57 \\
        10 & & & & i-Vector & & 21.48 & 48.32 & 17.42 & 37.79 & 34.71 \\
        11 & & & & x-Vector & & 20.83 & 48.53 & 32.29 & 43.10 & 35.88 \\
        12 & & & & SBE & & 20.44$^\dag$ & 47.70 & 17.31 & \textbf{36.11$^\dag$} & \textbf{33.76$^\dag$} \\
        \hline\hline
    \end{tabular}}
\end{table}

\begin{table}[ht]
    \caption{Performance comparison between the proposed spectral basis embedding feature based adaptation against i-Vector, x-Vector and LHUC adaptation on the \textbf{JCCOCC MoCA} corpus development (Dev) and evaluation (Eval) sets containing elderly speakers only. ``$18$M'' and ``$53$M'' refer to the number of model parameters. ``SBE'' denote spectral basis embedding features. $\dag$ denotes a statistically significant improvement ($\alpha=0.05$) is obtained over the comparable baseline i-Vector adapted systems (Sys. 2, 6 and 10).}
  \label{tab:recog-result-JM}
  \centering
  \renewcommand\arraystretch{1.0}
  \renewcommand\tabcolsep{2.0pt}
  \scalebox{0.88}{\begin{tabular}{c|c|c|c|c|c|c|c|c}
    \hline\hline
    \multirow{2}{*}{Sys.} &
    \multirow{2}{*}{\tabincell{c}{Model\\(\# Para.)}} &
    \multirow{2}{*}{\tabincell{c}{Data\\Aug.}} &
    \multirow{2}{*}{\# Hrs} &
    \multirow{2}{*}{\tabincell{c}{Adapt.\\Feat.}} & 
    \multirow{2}{*}{\tabincell{c}{LHUC}} &
    \multicolumn{3}{c}{CER\%} \\
    \cline{7-9} 
     & & & & & & Dev & Eval & O.V. \\ 
    \cline{7-8} 
    \hline\hline
    1 & \multirow{8}{*}{\tabincell{c}{Hybrid\\TDNN\\(18M)}} & \multirow{8}{*}{\cmark} & \multirow{8}{*}{156.9} & \xmark & \multirow{4}{*}{\xmark} & 26.87 & 23.71 & 25.28 \\
    2 & & & & i-Vector & & 25.46 & 22.80 & 24.12 \\
    3 & & & & x-Vector & & 25.06 & 21.93 & 23.49 \\
    4 & & & & SBE  & & \textbf{24.43$^\dag$} & \textbf{21.68$^\dag$} & \textbf{23.05$^\dag$} \\
    \cline{1-1}\cline{5-9}
    5 & & & & \xmark & \multirow{4}{*}{\cmark} & 25.77 & 22.94 & 24.35 \\
    6 & & & & i-Vector & & 24.73 & 22.12 & 23.42 \\
    7 & & & & x-Vector & & 24.18 & 21.48 & 22.82 \\
    8 & & & & SBE & & \textbf{23.59$^\dag$} & 21.42 & \textbf{22.50$^\dag$} \\
    \hline\hline
    9 & \multirow{4}{*}{\tabincell{c}{Conformer\\(53M)}} & \multirow{4}{*}{\cmark}  & \multirow{4}{*}{156.9} & \xmark & \multirow{4}{*}{\xmark} & 33.08 & 31.24 & 32.15 \\
    10 & & & & i-Vector & & 33.76 & 31.83 & 32.79 \\
    11 & & & & x-Vector & & 33.79 & 32.25 & 33.02 \\
    12 & & & & SBE & & \textbf{32.08$^\dag$} & \textbf{30.75$^\dag$} & \textbf{31.41$^\dag$} \\
    \hline\hline
  \end{tabular}}
\end{table}

Table~\ref{tab:recog-result-JM} shows the performance comparison of our proposed spectral basis embedding feature based adaptation against i-Vector, x-Vector and LHUC adaptation on the JCCOCC MoCA~\cite{xu2021speaker} data. Trends similar to those previously found on the DementiaBank Pitt corpus in Table~\ref{tab:recog-result-DBANK} can be observed in Table~\ref{tab:recog-result-JM}. Compared with the i-Vector adapted systems, a statistically significant overall WER reduction by up to $1.07\%$ absolute ($4.44\%$ relative) (Sys.4 v.s. Sys.2) can be obtained using the spectral embedding feature adapted hybrid TDNN systems. The SBE adapted E2E Conformer system outperformed its i-Vector baseline statistically significantly by $1.38\%$ absolute ($4.21\%$ relative) (Sys.12 v.s. Sys.10). 

\section{Discussion and Conclusions}
\label{sec-conclusion}
This paper proposes novel spectro-temporal deep feature based speaker adaptation approaches for dysarthric and elderly speech recognition. Experiments were conducted on two dysarthric and two elderly speech datasets including the English UASpeech and TORGO dysarthric speech corpora as well as the English DementiaBank Pitt and Cantonese JCCOCC MoCA elderly speech datasets. The best performing spectral basis embedding feature adapted hybrid DNN/TDNN and end-to-end Conformer based ASR systems consistently outperformed their comparable baselines using i-Vector and x-Vector adaptation across all four tasks covering both English and Cantonese. Experimental results suggest the proposed spectro-temporal deep feature based adaptation approaches can effectively capture speaker-level variability attributed to speech pathology severity and age, and facilitate more powerful personalized adaptation of ASR systems to cater for the needs of dysarthric and elderly users. Future researches will focus on fast, on-the-fly speaker adaptation using spectro-temporal deep features.

\section*{Acknowledgment}
This research is supported by Hong Kong Research Grants Council GRF grant No. 14200218, 14200220, Theme based Research Scheme T45-407/19N, Innovation \& Technology Fund grant No. ITS/254/19, PiH/350/20 and Shun Hing Institute of Advanced Engineering grant No. MMT-p1-19.

\ifCLASSOPTIONcaptionsoff
  \newpage
\fi

\bibliographystyle{IEEEtran}
\bibliography{main}

\end{document}